\documentclass[aps,twocolumn,showpacs,showkeys,preprintnumbers,superscriptaddress,nobibnotes,floatfix,longbibliography,notitlepage,nofootinbib]{revtex4-2}

\usepackage{graphicx}% Include figure files
\usepackage{dcolumn}% Align table columns on decimal point
\usepackage{bm}% bold math
\usepackage[dvipsnames]{xcolor}
\usepackage{enumitem}
\usepackage{amsmath}

\begin{document}

\preprint{LA-UR-25-26474}

\title{Measurement of the Liquid Argon Scintillation Pulse Shape Using Differentiable Simulation in the Coherent CAPTAIN-Mills Experiment}
\affiliation{Bartoszek~Engineering,~Aurora,~IL~60506,~USA}
\affiliation{Columbia~University,~New~York,~NY~10027,~USA}
\affiliation{University~of~Edinburgh,~Edinburgh,~United~Kingdom}
\affiliation{Embry$-$Riddle~Aeronautical~University,~Prescott,~AZ~86301,~USA }
\affiliation{University~of~Florida,~Gainesville,~FL~32611,~USA}
\affiliation{Los~Alamos~National~Laboratory,~Los~Alamos,~NM~87545,~USA}
\affiliation{Massachusetts~Institute~of~Technology,~Cambridge,~MA~02139,~USA}
\affiliation{Universidad~Nacional~Aut\'{o}noma~de~M\'{e}xico,~CDMX~04510,~M\'{e}xico}
\affiliation{University~of~New~Mexico,~Albuquerque,~NM~87131,~USA}
\affiliation{New~Mexico~State~University,~Las~Cruces,~NM~88003,~USA}
\affiliation{Texas~A$\&$M~University,~College~Station,~TX~77843,~USA}

\author{A.A.~Aguilar-Arevalo}
\affiliation{Universidad~Nacional~Aut\'{o}noma~de~M\'{e}xico,~CDMX~04510,~M\'{e}xico}
\author{S.~Biedron}
\affiliation{Element~Aero,~San~Leandro,~CA~94577,~USA}
\author{J.~Boissevain}
\affiliation{Bartoszek~Engineering,~Aurora,~IL~60506,~USA}
\author{M.~Borrego}
\affiliation{Los~Alamos~National~Laboratory,~Los~Alamos,~NM~87545,~USA}
\author{L.~Bugel\textsuperscript{\dag}}
\affiliation{Massachusetts~Institute~of~Technology,~Cambridge,~MA~02139,~USA}
\author{M.~Chavez-Estrada}
\affiliation{Universidad~Nacional~Aut\'{o}noma~de~M\'{e}xico,~CDMX~04510,~M\'{e}xico}
\author{J.M.~Conrad}
\affiliation{Massachusetts~Institute~of~Technology,~Cambridge,~MA~02139,~USA}
\author{R.L.~Cooper}
\affiliation{Los~Alamos~National~Laboratory,~Los~Alamos,~NM~87545,~USA}
\affiliation{New~Mexico~State~University,~Las~Cruces,~NM~88003,~USA}
\author{J.R.~Distel}
\affiliation{Los~Alamos~National~Laboratory,~Los~Alamos,~NM~87545,~USA}
\author{J.C.~D’Olivo}
\affiliation{Universidad~Nacional~Aut\'{o}noma~de~M\'{e}xico,~CDMX~04510,~M\'{e}xico}
\author{E.~Dunton}
\affiliation{Los~Alamos~National~Laboratory,~Los~Alamos,~NM~87545,~USA}
\author{B.~Dutta}
\affiliation{Texas~A$\&$M~University,~College~Station,~TX~77843,~USA}
\author{D.E.~Fields}
\affiliation{University~of~New~Mexico,~Albuquerque,~NM~87131,~USA}
\author{M.~Gold}
\affiliation{University~of~New~Mexico,~Albuquerque,~NM~87131,~USA}
\author{E.~Guardincerri}
\affiliation{Los~Alamos~National~Laboratory,~Los~Alamos,~NM~87545,~USA}
\author{E.C.~Huang}
\affiliation{Los~Alamos~National~Laboratory,~Los~Alamos,~NM~87545,~USA}
\author{N.~Kamp}
\affiliation{Massachusetts~Institute~of~Technology,~Cambridge,~MA~02139,~USA}
\author{D.~Kim}
\affiliation{University of South Dakota, Vermillion, SD 57069, USA}
%\affiliation{Texas~A$\&$M~University,~College~Station,~TX~77843,~USA}
\author{K.~Knickerbocker}
\affiliation{Los~Alamos~National~Laboratory,~Los~Alamos,~NM~87545,~USA}
\author{W.C.~Louis}
\affiliation{Los~Alamos~National~Laboratory,~Los~Alamos,~NM~87545,~USA}
\author{C.F.~Macias-Acevedo}
\affiliation{Universidad~Nacional~Aut\'{o}noma~de~M\'{e}xico,~CDMX~04510,~M\'{e}xico}
\author{R.~Mahapatra}
\affiliation{Texas~A$\&$M~University,~College~Station,~TX~77843,~USA}
\author{J.~Mezzetti}
\affiliation{University~of~Florida,~Gainesville,~FL~32611,~USA}
\author{J.~Mirabal}
\affiliation{Los~Alamos~National~Laboratory,~Los~Alamos,~NM~87545,~USA}
\author{M.J.~Mocko}
\affiliation{Los~Alamos~National~Laboratory,~Los~Alamos,~NM~87545,~USA}
\author{D.A.~Newmark}\email{Contact author: dnewmark@mit.edu}
\affiliation{Massachusetts~Institute~of~Technology,~Cambridge,~MA~02139,~USA}
\author{P.~deNiverville}
\affiliation{Los~Alamos~National~Laboratory,~Los~Alamos,~NM~87545,~USA}
\author{C.~O’Connor}
\affiliation{University~of~Florida,~Gainesville,~FL~32611,~USA}
\author{V.~Pandey}
\affiliation{Fermi National Accelerator Laboratory, Batavia, Illinois 60510, USA}
\author{D.~Poulson}
\affiliation{Los~Alamos~National~Laboratory,~Los~Alamos,~NM~87545,~USA}
\author{H.~Ray}
\affiliation{University~of~Florida,~Gainesville,~FL~32611,~USA}
\author{E.~Renner}
\affiliation{Los~Alamos~National~Laboratory,~Los~Alamos,~NM~87545,~USA}
\author{T.J.~Schaub}
\affiliation{University~of~New~Mexico,~Albuquerque,~NM~87131,~USA}
\author{A.~Schneider}
\affiliation{Los~Alamos~National~Laboratory,~Los~Alamos,~NM~87545,~USA}
\author{M.H.~Shaevitz}
\affiliation{Columbia~University,~New~York,~NY~10027,~USA}
\author{D.~Smith}
\affiliation{Embry$-$Riddle~Aeronautical~University,~Prescott,~AZ~86301,~USA }
\author{W.~Sondheim}
\affiliation{Los~Alamos~National~Laboratory,~Los~Alamos,~NM~87545,~USA}
\author{A.M.~Szelc}
\affiliation{University~of~Edinburgh,~Edinburgh,~United~Kingdom}
\author{C.~Taylor}
\affiliation{Los~Alamos~National~Laboratory,~Los~Alamos,~NM~87545,~USA}
\author{A.~Thompson}
\affiliation{Northwestern~University,~Evanston,~IL~60208,~USA}
\author{W.H.~Thompson}
\affiliation{Los~Alamos~National~Laboratory,~Los~Alamos,~NM~87545,~USA}
\author{M.~Tripathi}
\affiliation{University~of~Florida,~Gainesville,~FL~32611,~USA}
\author{R.T.~Thornton}
\affiliation{Los~Alamos~National~Laboratory,~Los~Alamos,~NM~87545,~USA}
\author{R.~Van~Berg}
\affiliation{Bartoszek~Engineering,~Aurora,~IL~60506,~USA}
\author{R.G.~Van~de~Water}
\affiliation{Los~Alamos~National~Laboratory,~Los~Alamos,~NM~87545,~USA}

\begingroup
\renewcommand\thefootnote{\dag}
\footnotetext{Deceased}
\endgroup

\collaboration{The CCM Collaboration}

\date{\today}

\begin{abstract}
The Coherent CAPTAIN-Mills (CCM) experiment is a liquid argon (LAr) light collection detector searching for MeV-scale neutrino and Beyond Standard Model physics signatures. Two hundred 8-inch photomultiplier tubes (PMTs) instrument the 7 ton fiducial volume with 50\% photocathode coverage to detect light produced by charged particles. CCM's light-based approach reduces requirements of LAr purity, compared to other detection technologies, such that sub-MeV particles can be reliably detected without additional LAr filtration and with $\mathcal{O}(1)$~parts-per-million of common contaminants. We present a measurement of LAr light production and propagation parameters, with uncertainties, obtained from a sample of MeV-scale electromagnetic events. The optimization of this high-dimensional parameter space was facilitated by a differentiable optical photon Monte-Carlo simulation, and detailed PMT response characterization. This result accurately predicts the timing and spatial distribution of light due to scintillation and Cherenkov emission in the detector. This is the first description of photon propagation in LAr to include several effects, including: anomalous dispersion of the index of refraction near the ultraviolet resonance, Mie scattering from impurities, and Cherenkov light production.
\end{abstract}

\maketitle

\section{\label{sec:intro}Introduction}
Liquid argon (LAr) is increasingly being used as a detection medium for large-scale neutrino detectors in the ton to kiloton range, and is an active area of detector technology research and development~\cite{DUNE:2024qgl,DarkSide-20k:2024yfq,SBND:2024vgn}. While single-phase LAr time projection chambers (TPCs) perform well in the $>$100 MeV energy range~\cite{Cavanna:2014iqa,Antonello:2013ypa,MicroBooNE:2016pwy,SBND:2020scp}, DEAP~\cite{DEAP:2025shk,DEAP:2014pug} and CCM~\cite{CCM:2021yzc,CCM:2021jmk,CCM:2021leg,CCM:2023itc,Tripathi:2024jnq,Dunton:2022dez} have demonstrated that optical LAr detectors performs exceptionally well for MeV-scale interactions. At this energy scale, low-energy weakly-interacting physics and a range of new physics scenarios can be targeted. Optical LAr detectors use a high-coverage of photomultiplier tubes (PMTs), or other photo-sensors, surrounding the bulk volume of LAr to observe the photons that are copiously produced when charged particles traverse the medium. This design is similar to scintillator-oil-based detectors, such as Borexino \cite{Borexino:2000uvj} and SNO+ \cite{SNO:2015wyx}. Despite the challenges associated with handling a cryogenic liquid, LAr-based optical detectors offer several key advantages--- such as high scintillation light yield per unit energy deposited, transparency to optical photons, and cryogenic conditions that can suppress dark rate currents in certain photo-sensors. These features enable event-by-event observation of Cherenkov light produced by sub-MeV electrons, as demonstrated in Ref.~\cite{companion_paper}.

This paper characterizes light production and propagation in the CCM experiment. CCM is unique in that, among LAr detectors that utilize light collection only, it has the largest volume and the highest photo-coverage. As such, CCM is an ideal test bed for studying LAr scintillation physics. An important aspect of the CCM design is that the LAr is not purified beyond the levels present upon delivery, which leaves parts-per-million (ppm) of nitrogen, oxygen, and water contaminants. These contaminant levels reduce the overall light yield, but drastically increase the absorption of free electrons to the point that TPC detectors cannot reliably detect particles. For CCM, this does not significantly impact our sensitivity to MeV-scale physics, but reduces complexity and cost of the detector design.

This is the first experimental validation of the wavelength-resolved index of refraction fit using a damped harmonic oscillator model in LAr, which describes the anomalous dispersion near the ultraviolet (UV) resonance~\cite{Rahman:2024zhp}. While previous fits for the index of refraction in LAr using the Sellmeier dispersion relation~\cite{born1999principles,Babicz:2020den,Grace:2015yta} are valid far from the UV resonance, they diverge to infinity at the resonance, which is in a crucial wavelength regime for both scintillation light and Cherenkov radiation. We use the damped harmonic oscillator fit framework described in Ref.~\cite{Rahman:2024zhp}, and allow variations in the UV regime where there is only a single reference measurement~\cite{Babicz:2020den}.

We also explore Mie scattering in the bulk LAr. While optical photons will only Rayleigh scatter off the argon atoms, due to their small size, our measurements demonstrate that optical photons may Mie scatter off of larger impurities in the LAr. This motivates further study of the optical properties, especially in larger detectors where photons travel much longer path lengths to detection.

To describe light production and propagation in CCM, we developed a differentiable simulation model to simultaneously constrain more than 20 parameters in a binned likelihood optimization. Differentiable simulations, which are commonly used in machine learning applications, ascribe a gradient with respect to physics parameters on the unit of measure~\cite{degrave2016differentiable,6386109,newbury2024reviewdifferentiablesimulators}. For photoelectrons (PEs) detected by PMTs in the simulation, one can track the originating photon information, such as emission wavelength, distance traveled, and wavelength shifts, as well as the wavelength and location upon detection. These properties can then be used to re-weight the probability of PE observation under a different physics scenario. To calculate the prediction as a function of absorption length, for example, the PE probability is re-weighted by the ratio of the exponential probability distribution functions for the simulated absorption length and the desired absorption length, given the distance traveled by the photon in LAr. In aggregate, this produces accurate expectation values without the need to re-simulate.

This approach is advantageous in two respects--- first, the re-weightability of the simulation significantly reduces computational time required to produce optical model predictions that require detailed Monte-Carlo modeling. Instead of simulating at each parameter set, which is prohibitive with more than 20 parameters, a single simulation set assuming nominal parameter values can be used to make predictions for a wide range of parameter space. Second, the analytic gradient information enables efficient gradient-based minimization~\cite{minimizer1,minimizer2}. This improves fit convergence and computation time, enabling optimization of a large set of parameters without compromising on the physics modeling. 

In order to measure the light production and propagation parameters, this study uses data from a $^{22}$Na calibration source and compares the time and spatial structure of light production in CCM to simulation. This source produces $\sim$MeV-scale gamma-rays that typically Compton scatter in the fiducial volume of LAr, creating electrons that both scintillate and, if sufficiently energetic, produce Cherenkov radiation. As described in Ref.~\cite{companion_paper}, the Cherenkov light component is most notable in the 6~ns region preceding the reconstructed event start times. This work, however, will demonstrate that Cherenkov light is a $>10\%$ component of the total expectation in the entire detector at times $t \leq 5$~ns relative to the reconstructed event start time $t=0$, necessitating descriptions of both the Cherenkov and scintillation signals simultaneously. Subsequent sections describe the elements of the simulation in detail and report measured values of the simulation parameters with uncertainties. The broad categories of parameters are those associated with scintillation light production, Cherenkov light production, scattering and absorption in the bulk material, wavelength shifting of UV to visible photons, and PMT timing response.

\section{\label{sec:lar_scint_physics}Liquid Argon Scintillation Physics}
Ionizing radiation in liquid argon can create self-trapped excitons, resulting in excited dimers Ar$_2^*$ that decay by producing a vacuum ultraviolet (VUV) photon~\cite{Doke:1990rza}. These excited dimers have two characteristic time constants from the singlet and triplet spin configurations~\cite{Whittington:2014aha,Segreto:2020qks,DEAP:2020hms}. The singlet state has a fast time constant of $\mathcal{O}(5{\rm ~ns})$ and the triplet state has a slower time constant of $\mathcal{O}(1000{\rm ~ns})$. In addition to these two time constants, there is experimental evidence for a so-called ``intermediate'' time component that affects the photon time distribution between 30~ns and 100~ns~\cite{Hofmann:2013vva,DEAP:2020hms}. This additional time structure is hypothesized to be caused by delayed electron recombination with the excited argon dimers. Eq.~\ref{eq:lar_pulse_shape} describes the normalized photon time distribution, where the first term corresponds to the singlet state decay, the second term corresponds to the triplet state decay, and the final term describes the intermediate time recombination component~\cite{Hofmann:2013vva,DEAP:2020hms}. $R_s$ and $R_t$ describe the proportion of scintillation photons produced by the singlet and triplet states, respectively. The characteristic decay time constants are $\tau_s$, $\tau_t$, and $\tau_{rec}$, associated with the singlet, triplet, and intermediate components of photon time distribution, respectively.

\begin{equation}
I(t) = \frac{R_s}{\tau_s}e^{-t / \tau_s} + \frac{R_t}{\tau_t}e^{-t / \tau_t} + \frac{1 - R_s - R_t}{(1 + t / \tau_{rec})^2~\tau_{rec}}
\label{eq:lar_pulse_shape}
\end{equation}

\begin{figure}[h]
  \centering
  \includegraphics[width=\linewidth]{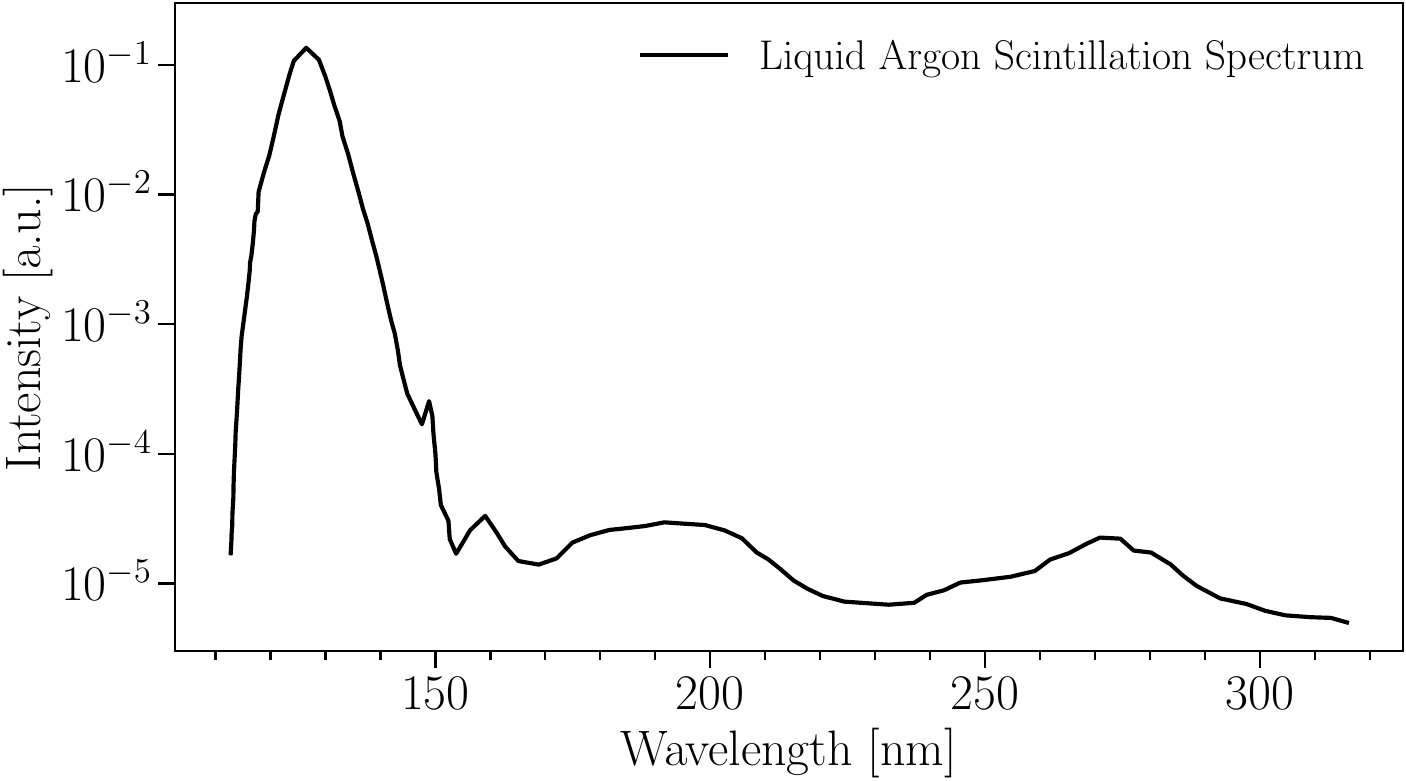}
  \caption{Wavelength dependence of liquid argon scintillation photon emission, as measured by Ref.~\cite{Heindl:2010zz}. The majority of photons are emitted at approximately 128~nm with a FWHM of approximately 8~nm due to the decay of the excited dimer $^1\Sigma$ state, or the so-called first excimer continuum. There is additional emission between 175~nm and 250~nm due to the third excimer continuum~\cite{PhysRevA.43.6089}. See Ref~\cite{Heindl:2010zz} for full discussion of the wavelength dependence of LAr scintillation.}
  \label{fig:lar_scint_spectrum}
\end{figure}

For both the singlet and triplet excited dimer states, Ar$_2^*$ scintillation photon emission is peaked around $128~\rm{nm}$ vacuum wavelength~\cite{Heindl:2010zz}. Fig.~\ref{fig:lar_scint_spectrum}, the digitized LAr scintillation emission spectrum from Ref.~\cite{Heindl:2010zz}, demonstrates the full wavelength dependence of scintillation emission. The majority of scintillation photons are emitted around $128~\rm{nm}$ with a full width at half maximum (FWHM) of approximately 8~nm. The weaker emission peaks are hypothesized to be due to the third excimer continuum and possible xenon contamination~\cite{PhysRevA.43.6089,Heindl:2010zz}. 

The VUV wavelength of scintillation photons requires wavelength shifters to allow detection using PMTs. CCM employs the commonly used wavelength shifter tetraphenyl butadiene (TPB), which efficiently absorbs VUV photons and re-emits into the visible spectrum, primarily between 400~nm and 550~nm~\cite{Benson:2017vbw,doi:10.1021/j100052a011}. The TPB is installed on the CCM detector walls and some PMT faces; thus, unlike oil-based scintillation detectors, wavelength-shifting happens at the edges of the detector rather than within the bulk. The benefits of this design for Cherenkov radiation separation are discussed in  Ref.~\cite{companion_paper} and are important to sub-MeV electron identification. 

The presence of impurities in LAr is known to affect the scintillation light time structure. Nitrogen and other contaminants reduce the lifetimes of the dimer states, especially the long-lived triplet state, by quenching~\cite{WArP:2008rgv,MicroBooNE:2022pcx}. This will affect the ratio of singlet to triplet states that contribute scintillation light~\cite{Mavrokoridis:2011wv}. Depending on the energy scale of the physics goals, the reduction in the long time-scale component of scintillation photon emission due to impurities can be leveraged to improve reconstruction. In a high rate environment where the physics goals necessitate reconstruction of multiple events in a limited time period, a shorter triplet lifetime can reduce backgrounds from late-emitted photons produced by previous particle interactions. 

\section{\label{sec:lar_cherenkov_physics}Cherenkov Light in Liquid Argon}
Cherenkov light is produced when a charged particle travels faster than the speed of light in a medium~\cite{Cherenkov:1937mnd}. The emitted photons propagate in a cone centered along the direction of particle motion and with an opening angle given by $\cos(\theta) = 1 / \beta n$, where $n$ is the index of refraction that depends on wavelength $\lambda$ of the emitted photon and $\beta$ is the particle velocity. If the emitted light does not scatter substantially before detection by the PMTs at the edges of the detector, the signal will form a disk, with the center aligned with the direction of particle travel and with radius related to the distance from the particle origin to the wall. As a result, the detection of Cherenkov light at the wall can be used to isolate particles that have velocities above the Cherenkov threshold, and the position and thickness of the ring can be used for track reconstruction~\cite{Patterson:2009ki,Super-Kamiokande:2019gzr}. Because multiple scattering of a particle depends on the particle mass, smearing of the ring edge can also allow for particle identification. 

The index of refraction, in addition to determining the group velocity of photons in a medium, is very important in Cherenkov radiation. The threshold for Cherenkov radiation and the number of Cherenkov photons, $N$, emitted per unit path length $dx$ depends on the energy of the particle and the index of refraction of the medium, as demonstrated by the Frank-Tamm formula, Eq.~\ref{eq:frank_tamm}~\cite{10.1093/ptep/ptac097}. This equation describes the relationship between the number of Cherenkov photons emitted per unit path length and per unit wavelength $\lambda$ as a function of the fine structure constant $\alpha$, the charge of the particle $z$, the velocity $\beta$, and the index of refraction $n(\lambda)$. In LAr, Sinnock and Smith measured the index of refraction at various wavelengths between 350~nm and 650~nm vacuum wavelengths~\cite{Sinnock:1969zz}. Additionally, measurement of the group velocity in liquid argon at 128~nm vacuum wavelength provides a data point for the index of refraction in the VUV range~\cite{Babicz:2020den}. 

\begin{equation}
\frac{d^2 N}{dx d\lambda} = \frac{2 \pi \alpha z^2}{\lambda^2} \left( 1 - \frac{1}{\beta^2 n^2(\lambda)} \right)
\label{eq:frank_tamm}
\end{equation}

Generally, the Sellmeier dispersion relationship describes the index of refraction as a function of wavelength for transparent media~\cite{born1999principles}. Eq.~\ref{eq:sellmeier} describes this relationship which can be fit to experimental data to extract the Sellmeier coefficients, $a_0$ and $a_i$ while $\lambda_i$ are absorption resonances~\cite{Grace:2015yta,Babicz:2020den}. While this method is appropriate far from the resonances, the index of refraction diverges as $\lambda \rightarrow \lambda_i$.

\begin{equation}
n^2(\lambda) = a_0 + \sum_i \frac{a_i \lambda^2}{\lambda^2 - \lambda_i^2}
\label{eq:sellmeier}
\end{equation}

In LAr, there is a resonance in the VUV range at 106.6~nm~\cite{Lane1968} and in the infrared range at 908.3~nm~\cite{Arai1978}. The index of refraction in the VUV range is very important for both scintillation physics and Cherenkov physics. For the approximately 128~nm scintillation photons, the index of refraction in the VUV range impacts the group velocity as well as other optical processes such as refraction. For Cherenkov radiation, which is only produced when $\beta n > 1$, the index of refraction determines both the energy threshold for particles to produce Cherenkov photons and the quantity of Cherenkov photons produced. If the Sellmeier dispersion relation is used, the total number of emitted photons is ill-defined because of the divergence near 106.6~nm, and in practice an arbitrary lower bound on $\lambda$ must be introduced to limit the number of photons to a finite quantity and this sets an energy threshold.

A more accurate treatment of the index of refraction accounts for the absorptive effects near the resonances. Ref.~\cite{Rahman:2024zhp} uses a damped harmonic oscillator model to simultaneously fit for the real index of refraction and imaginary absorption coefficient. This results in a wavelength dependent index of refraction that is well-behaved across resonances, described by Eq.~\ref{eq:ho_rindx}. The variables $a_0$ and $a_{UV}$ are similar to the Sellmeier coefficients, while $\gamma_{UV}$ describes the damping at the UV resonance. Eq.~\ref{eq:ho_rindx} is fit to the existing experimental data of the index of refraction in LAr~\cite{Sinnock:1969zz,Babicz:2020den}, resulting in best fit parameters are listed in Table~\ref{table:ho_rindex_values}~\cite{Rahman:2024zhp}.

\begin{equation}
n = a_0 + a_{UV} \left ( \frac{\lambda_{UV}^{-2} - \lambda^{-2}}{ (\lambda_{UV}^{-2} - \lambda^{-2})^2 + \gamma_{UV}^2 \lambda^{-2}} \right )
\label{eq:ho_rindx}
\end{equation}

\begin{table}[]
  \caption{Damped harmonic oscillator model index of refraction fit parameters as described by Ref.~\cite{Rahman:2024zhp}. See Eq.~\ref{eq:ho_rindx} for full implementation of this model.}
  \begin{ruledtabular}
    \begin{tabular}{cc}
      Parameter & Fit Value  \\
      \hline
      $a_0$ & $1.10232$  \\
      $a_{UV}$ & $0.00001058~\rm{nm}^{-2}$  \\
      $\gamma_{UV}$ & $0.002524~\rm{nm}^{-1}$ \\
    \end{tabular}
  \end{ruledtabular}
  \label{table:ho_rindex_values} 
\end{table}

\begin{figure}[h]
  \centering
  \includegraphics[width=\linewidth]{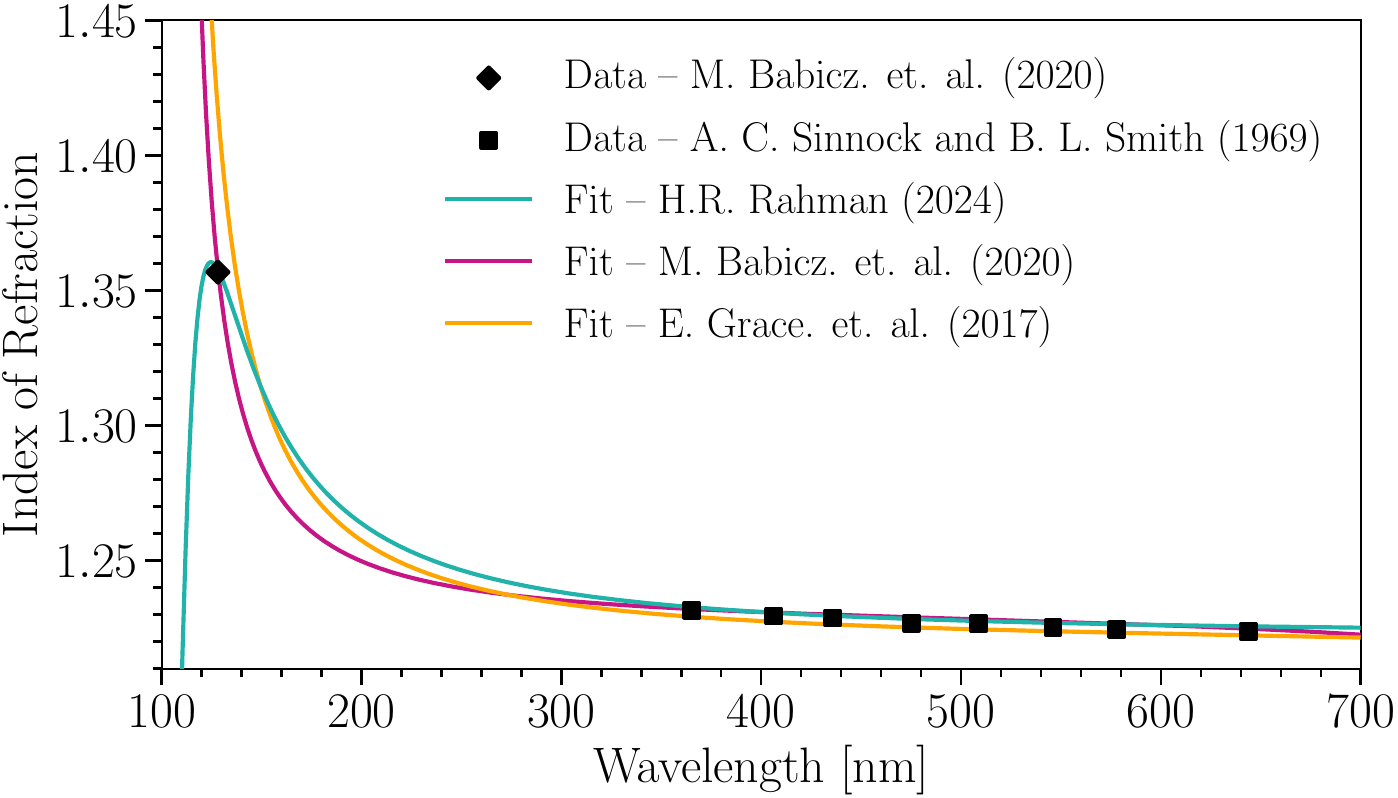}
  \caption{Comparison of literature values of the index of refraction in LAr. The single data point of the index of refraction measured using LAr scintillation light, at $\sim$128~nm, is denoted by the diamond marker~\cite{Babicz:2020den}. Data taken in the visible range, 350~nm to 650~nm, are indicated by the square markers~\cite{Sinnock:1969zz}. There are three different fits to the data -- the orange line fits the Sellmeier dispersion relation to the data points in the visible~\cite{Grace:2015yta}. The pink line fits the Sellmeier dispersion as well but includes all the data points~\cite{Babicz:2020den}. The blue line is a fit using the damped harmonic oscillator parametrization of the index of refraction to all available data points~\cite{Rahman:2024zhp}.}
  \label{fig:rindex}
\end{figure}

Fig.~\ref{fig:rindex} shows the experimental data points, indicated by the diamond and box markers~\cite{Babicz:2020den,Sinnock:1969zz}, the fits using the Sellmeier dispersion relation (M. Babicz et. al.~\cite{Babicz:2020den} and E. Grace et. al.~\cite{Grace:2015yta}), and the fit using a damped harmonic oscillator model that accounts for the imaginary absorption term (H.R. Rahman~\cite{Rahman:2024zhp}). The harmonic oscillator fit has an anomalous dispersion portion to the index of refraction below 128~nm due to the inclusion of absorption physics. These fits also differ in the transition in the index of refraction between the VUV and the visible wavelength regions. Additional measurements of the index of refraction below 350~nm in LAr would greatly improve this wavelength resolved description.

\section{\label{sec:detector} Elements of the CCM Experiment}

This section describes the environment in which CCM is deployed and the design of the detector. Here, we also discuss the impact of the choice to not install a filtration system.

\subsection{\label{sec:CCMenviro} CCM at the Lujan Spallation Neutron Center}

The CCM200 detector is located at the Lujan Center on the Los Alamos Neutron Science Center (LANSCE) beamline. The Lujan neutron spallation facility provides an 800 MeV proton beam that impinges on a tungsten target. The detector is located 90$^\circ$ off axis and 23~m away from the target, behind 6~m of steel, 3.5~m of concrete, and 5~cm of borated polyethylene shielding. This positioning and shielding affords maximum sensitivity to neutrinos from pion decay-at-rest and Beyond Standard Model (BSM) particles coupling to pions, electrons, and photons within the beam dump. The proton bunches are pulsed at 20~Hz with approximately 100~$\mu$A current in a 145~ns FWHM triangular beam spill window. This short beam spill allows for timing cuts that isolate signals produced from neutrinos and nearly-speed-of-light BSM particles, while rejecting the slower-moving neutron backgrounds that penetrate the shielding. The pulsed beam also reduces steady-state random backgrounds. Further discussion of the beam is provided in Ref.~\cite{CCM:2021leg}.

\subsection{\label{sec:ccmdesign} The Detector Design}
CCM200, a light-based detector, employs a 2.58 m diameter $\times$ 2.25 m tall upright cylindrical cryostat~\cite{Tripathi:2024jnq,Dunton:2022dez}. This cryostat is instrumented with 200 8-inch R5912-Y002 10 stage cryogenic Hamamatsu PMTs, providing 50\% photocoverage. To increase total visible light collection, the walls of the detector are coated in Mylar reflective foils with a 2.8~$\mu$m thick layer of evaporatively-coated TPB that wavelength shifts VUV and UV photons into the visible spectrum for detection. Additionally, 80\% of the PMTs are evaporatively-coated with a 2.0~$\mu$m layer of TPB. The remaining 20\% of the PMTs are uncoated, an aspect of the design that is important for Cherenkov light separation, as discussed in Ref.~\cite{companion_paper}.

\begin{figure}[h]
  \centering
  \includegraphics[width=\linewidth]{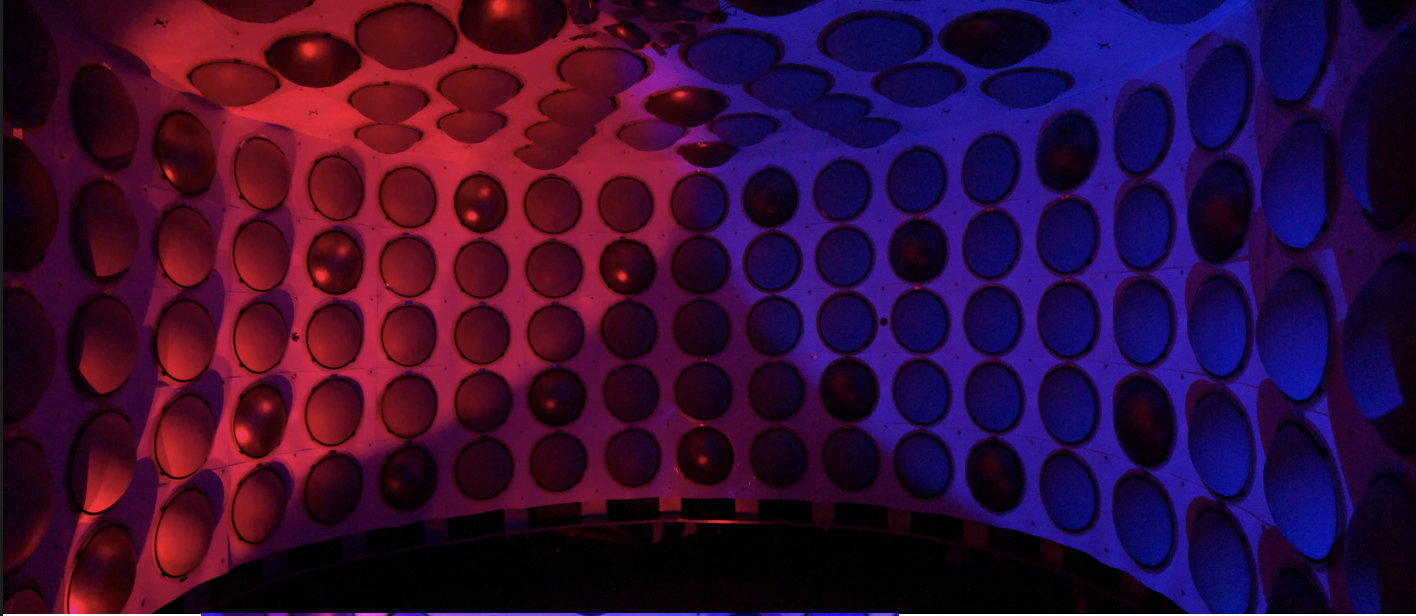}
  \caption{Interior view of the CCM200 detector during assembly. The fiducial region, enclosed by the PMTs and support structure, encompasses 7~tons of LAr. The PMTs protrude 6.2~mm into the fiducial volume, providing full photocathode exposure. The reflective PMTs do not have wavelength shifter coating while the matte PMTs are coated in TPB. 20\% of the PMTs are uncoated for a total of 40 uncoated PMTs.}
  \label{fig:ccm200_detector}
\end{figure}

Fig.~\ref{fig:ccm200_detector} shows the interior of the fiducial region of the detector during assembly. The uncoated tubes can be identified as they are more reflective than the PMTs coated in TPB, which scatters the visible light. The PMT signals are digitized and read out using CAEN V1730/V1730S 500 MHz boards, providing digitization in 2~ns bins. This sets the timing resolution as the spread in the PMT electron transit times are $\mathcal{O}({1\rm ~ns})$ and the FWHM of the a typical single PE signature is approximately $6$~ns for most PMTs.

\subsection{\label{sec:impurities} Commercial Argon Without Filtration}
An unusual aspect of the CCM design is that the detector is run with commercially supplied argon that is not further filtered.  The argon is allowed to boil off continuously with periodic replacement at the rate of $\sim$380~liters per day out of a total volume of 7000~liters.

CCM200 is filled with LAr from atmospheric sources, provided by Matheson Tri Gas, Albuquerque, New Mexico. The manufacturer's maximum impurity specifications are 1.95~ppm oxygen, 2.50~ppm nitrogen, and 0.01~ppm water contaminant levels. While taking data, we continuously monitor detector conditions using commercially available oxygen and nitrogen analyzers. These monitors measure the impurity levels at $2.2\pm0.5$~ppm oxygen and $0.1\pm0.1$~ppm nitrogen, confirming the impurity levels are relatively stable over the course of data collection, enabling robust characterization of an optical model. 

Impurities quench scintillation through collisions with the excited argon dimers and can absorb emitted scintillation photons. Cryogenic filtration systems, however, are costly, complex and do not remove nitrogen~\cite{WArP:2008dyo,WArP:2008rgv,Jones:2013bca}. Because LAr scintillation produces an average of 40,000 photons per MeV of deposited kinetic energy~\cite{Doke:1990rza}---a very high rate that allows for high photon loss without compromising the physics goals---we have chosen to run CCM200 without a filtration system.

In fact, low levels of impurities that quench UV light aid in separation of visible Cherenkov photons from scintillation signals. This is because the impurities absorb primarily in the VUV and UV wavelength regimes, while the clearest indication of Cherenkov light is obtained from visible wavelengths. Prompt observation of photons by the uncoated PMTs is used to preferentially select for visible Cherenkov light against a larger scintillation background~\cite{companion_paper}, so allowing impurities to reduce the observed scintillation photons improves Cherenkov light separation.

\section{\label{sec:data_collection}Photoelectron Pulse Reconstruction and Data Collection}
From the digitized voltage changes of the PMT waveforms  as a function of time, the amplitude and time structure of PE pulses are reconstructed and used as the basic unit of data. CCM utilizes the Lawson-Hanson NNLS algorithm for PE pulse unfolding following the methods developed by the IceCube Collaboration~\cite{lawson1974least,IceCube:2013dkx}.

\begin{figure}[h]
  \centering
  \includegraphics[width=\linewidth]{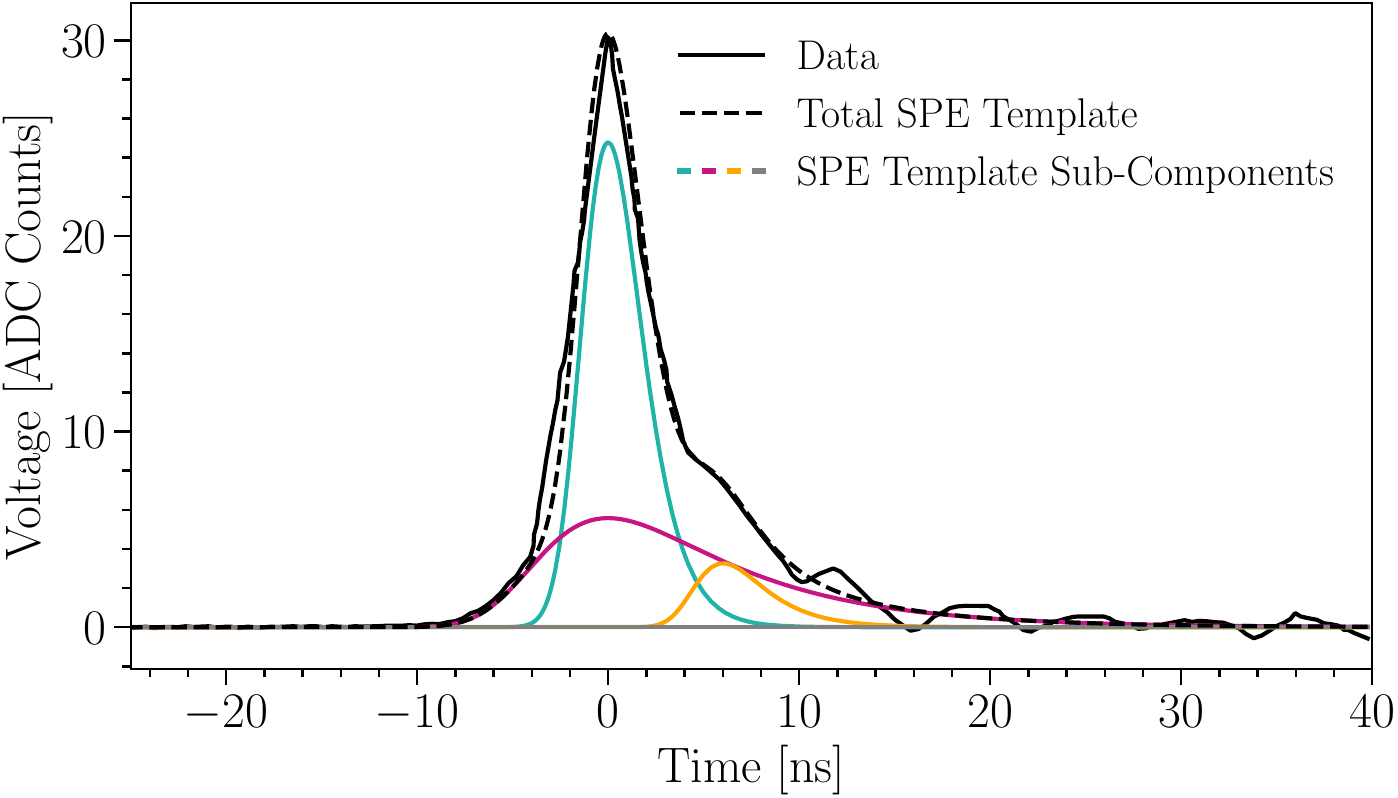}
  \caption{Example SPE template used for pulse unfolding for a typical PMT. The data (black solid line) is the average SPE pulse shape. The total SPE template fit used for pulse unfolding is the black dashed line and the four sub-components are the blue, magenta, orange, and gray lines.}
  \label{fig:spe_template}
\end{figure}

The average single PE (SPE) shape as a function of time is characterized for every PMT in CCM and employed for this template-based method of PE pulse unfolding. Fig.~\ref{fig:spe_template} is an example of this template for a typical PMT. The data represents the average of 8,750,705 single pulses identified in digitized waveforms collected across the 2022 run period. Pulses were selected using a derivative-based algorithm that looks for a sequence of positive, negative, and then positive derivatives--- corresponding to a rising edge, a falling edge, and a second rise of the pulse based on observed overshoot after single pulses~\cite{Kaptanoglu:2017jxo}.

The averaged data is then fit using an SPE template, derived from the functional form originally developed by the IceCube Collaboration~\cite{I3DOMCalibration,IceCube:2008qbc,IceCube:2016zyt}, detailed in Eq.~\ref{eq:spe_template}. The variable $h$ determines the height and the variables $b_1$ and $b_2$ vary the shape of the pulse. We use a combination of four SPE templates (blue, magenta, orange, and gray lines in Fig.~\ref{fig:spe_template}) to fully describe the overall characteristic shape of an SPE (black dashed line in Fig.~\ref{fig:spe_template}) for every PMT individually to capture the time dependence. One notable feature of the PMT characteristic pulse shape is the second peak contribution around time $t=6$~ns relative to the peak time, which has been observed previously for similar models of PMTs~\cite{Caldwell:2013oea}. Two of the SPE templates account for the main peak at $t=0$ and secondary peak at $t=6$~ns and the other two templates help describe the width of the characteristic SPE shape. After this template is used for pulse unfolding, the distributions of reconstructed pulse amplitudes are characterized and calibrated such that the Gaussian mean defines 1~PE.

\begin{equation}
\begin{aligned}
w(t) &= \frac{c}{\left(e^{-(t - t_0)/b_1} + e^{(t - t_0)/b_2}\right)^8}; \\
c &= \frac{h}{
  b_1^{(8b_1)/(b_1 + b_2)} \cdot
  b_2^{(8b_2)(b_1 + b_2)} /
  (b_1 + b_2)^8
}
\end{aligned}
\label{eq:spe_template}
\end{equation}

The calibration measurements that follow use data from a 3 $\mu$Ci $^{22}$Na source, which was taken while the beam was off. $^{22}$Na has two primary decay channels, $\beta^+$ and electron capture~\cite{BASUNIA201569}. The $\beta^+$ decay channel has approximately 90\% branching ratio and produces a 0.546~MeV kinetic energy positron (which annihilates producing two back-to-back 0.511~MeV gamma-rays) and a 1.275~MeV gamma-ray from subsequent $^{22}\rm{Ne}^*$ de-excitation. The electron capture decay channel has approximately 10\% branching ratio and produces only a single 1.275~MeV gamma-ray from $^{22}\rm{Ne}^*$ de-excitation. These gamma-rays then typically Compton scatter in the LAr, producing approximately 1~MeV and sub-MeV electrons, which scintillate and can create Cherenkov radiation.

The source is encapsulated in approximately 1~mm of stainless steel, inserted on a stainless steel bayonet through a central flange at the top of the detector, and positioned at the midline of the detector. Data is collected for a 16~$\mu$s data acquisition window using an external 20~Hz trigger. Events are identified using a charge over threshold algorithm. We require a charge threshold of at least 20~PE in the entire detector in a time window of 20~ns to define an event, then use a finer grained event finder requiring 3~PE in a 2~ns time window to define the start time of that event. In order to maintain purity of events, the following data quality cuts are applied.

\begin{enumerate}[align=left, label=\arabic*., labelwidth=0pt, labelsep=0.2em, leftmargin=0.0em, itemindent=0pt]
    \item Cosmic muon---  PE pulse series with cosmic muons, defined as greater than 200~PE in 2~ns, identified anywhere in the digitization window are eliminated entirely. Since high charge muons typically have associated long-time scale PE pulses from triplet scintillation light production, or, in the case of muon decay, have correlated Michel electron emission, we remove all events in the same trigger window to maintain purity of sodium data selection.
    
    \item Surrounding event--- Events must have no other events identified in the preceding and following 2.2~$\mu$s. This is designed to avoid contamination from the $\mathcal{O}(\rm{\mu s})$ scale triplet lifetime of LAr excited dimers. It additionally allows for accumulation of PE pulse series out to 2~$\mu$s from the event start time for fitting with limited contamination from background or overlapping sodium events.
    
    \item Radius--- Positions are roughly reconstructed using the first 20~ns of PE pulse series data by calculating a charge weighted average of PMT locations. The radius cut requires the reconstructed radius to be $\leq25$~cm relative to the origin, isolating events in the center of the detector where the source is located. This cut was designed to eliminate random backgrounds, which typically occur in the outer volume of the fiducial region. 
    
\end{enumerate}

\begin{figure}[h]
  \centering
  \includegraphics[width=\linewidth]{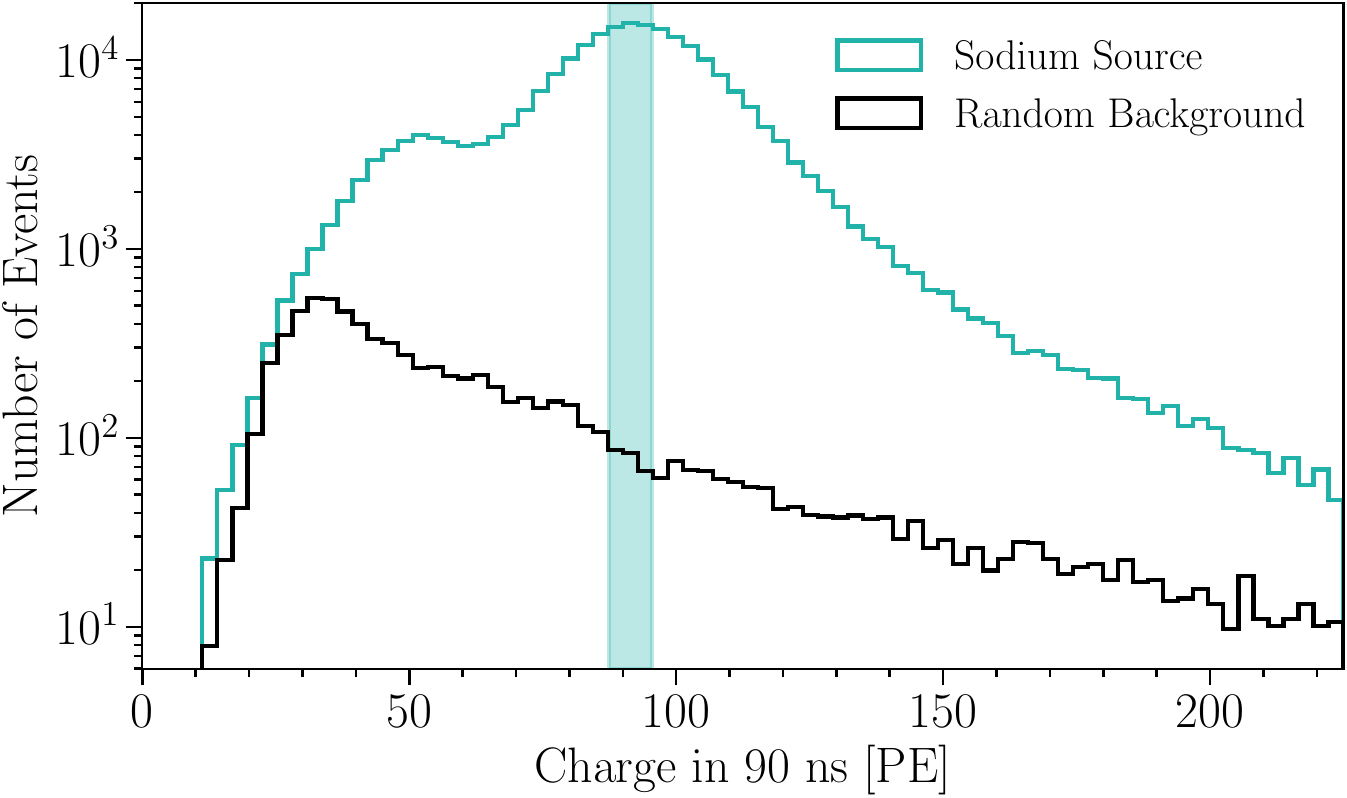}
  \caption{Distribution of total charge in the first 90~ns of an event. After data quality cuts are applied, described in Sec.~\ref{sec:data_collection}, charge in the first 90~ns of an event is used as a proxy for energy. For data collected with the sodium source inserted (blue) two clear peaks can be identified. The peak around 50~PE is due to the electron capture decay pathway for the sodium source and the higher charge peak at 100~PE is from the $\beta^+$ decay pathway. Data collected when the sodium source was removed (black) shows the exponential backgrounds due to ambient radioactivity. For fitting the optical model parameters, data within $\pm$4~PE (shaded band) of the high charge peak was selected and accumulated.}
  \label{fig:sodium_charge}
\end{figure}

These cuts ensure that we can collect events over a long duration with a high purity of $^{22}$Na data on every PMT. Fig~\ref{fig:sodium_charge} shows the distribution of charge in the first 90~ns of an event after applying the above data quality cuts. The blue line is using data collected while the sodium source was inserted and the black line is data collected without the source present, referred to as ``random background". The two decay pathways for the sodium source, electron capture and $\beta^+$ decay, can be easily identified at $\sim$50~PE and $\sim$100~PE respectively. For fitting the light profile and other optical model parameters, we selected events with charges $\pm$4~PE of the $\beta^+$ decay peak, reducing the random background event contamination rate to $\leq0.52\%$. After selecting events, PE pulse series were accumulated on every PMT using global event start times and the accumulated data was used for fitting.

\section{\label{sec:fitting_strategy}Fitting Strategy}
We used the Monte Carlo simulation package~\texttt{Geant4} to produce the simulation sets used in optical model calibration~\cite{GEANT4:2002zbu}. The geometry of the detector and radioactive source are fully described in the simulation, and the selected physics models are drawn from the~\texttt{FTFP\_BERT\_HP} physics list with a few notable exceptions. General electromagnetic physics is modeled using the~\texttt{G4EmStandardPhysics\_option4}, but the Compton scattering process is replaced with the~\texttt{G4PenelopeComptonModel} to ensure accurate modeling at relevant photon energies~\cite{GEANT4:2002zbu}. Default optical physics processes are enabled through~\texttt{G4OpticalPhysics}, but a modified copy of~\texttt{G4Cerenkov} is used to account for anomalous dispersion in the LAr index of refraction.

Optical photon simulation is often computationally costly and optimizing a high-dimensional parameter space requires producing simulation predictions at a large number of parameter-space points. To reduce the computational cost of optimization without compromising the physics modeling, we augment the~\texttt{Geant4} photon propagation to create a differentiable and re-weightable simulation. 

For every photon created in the simulation, we track the properties and interactions that affect or would be affected by the physical processes we wish to modify in our parameter optimization. These include the photon's distance traveled in LAr before and after wavelength shifting, wavelength at production and detection, originating process (namely scintillation or Cherenkov), time delay from scintillation physics, and time delay from PMT post-pulsing among others. We then produce a simulation set of $^{22}$Na decay events using a nominal set of scintillation time constants, bulk LAr properties, and other parameters, that is processed in the same manner as data with event identification, selection cuts, and time alignment.

To calculate the expectation for another point in the parameter space, every simulated PE is individually re-weighted according to a ratio of probability distribution functions evaluated at the nominal and new parameter values using forward automatic differentiation~\cite{icecube_phystools}. The detailed information recorded about each photon allows this re-weighting to account for timing and spatial dependencies that result from differences in photon creation, propagation, and detection. Below is a brief overview of the main parameters varied in the fit and their treatment in the re-weighting procedure.

\begin{enumerate}[align=left, label=\arabic*., labelwidth=0pt, labelsep=0.2em, leftmargin=0.0em, itemindent=0pt]

\item Scintillation Pulse Shape--- The main contribution to the time structure of the pulse series is the scintillation physics, described in Sec.~\ref{sec:lar_scint_physics}. Photons are re-weighted analytically using the ratio of the desired and nominal pulse shapes $I(t, R_s, R_t, \tau_s, \tau_t, \tau_{rec})$, Eq.~\ref{eq:lar_pulse_shape}. The scintillation pulse shape was set to unity for Cherenkov photons. 

\item Photon Absorption--- Wavelength resolved absorption due to contaminants is among the least constrained parameters in this model because CCM200 is unique in not filtering the LAr beyond that of delivery. By tracking photon wavelengths and distances traveled (including separate record-keeping of photons as they were created in the LAr and after wavelength shifting), photon absorption is re-weighted analytically using the ratio of exponential decay PDFs; $w(d, x(\lambda)) = e^{-d / x(\lambda)}$ for the distance traveled $d$ and the wavelength resolved absorption length $x(\lambda)$. 

\item Index of Refraction--- The index of refraction plays a central role in determining the optical properties of the medium, including characteristics of Cherenkov radiation. Because many aspects of optical photon transport depend non-linearly on the index of refraction, it is not straightforward to analytically re-weight between different models. To address this, simulations are performed at discrete points in the parameter space, and the results are interpolated to estimate how adjustments to the damped harmonic oscillator model of the index of refraction affect the expectation.

\item Scattering Lengths--- This work explores Rayleigh and Mie scattering in the bulk LAr. Directly modeling the impact of varying scattering lengths on photon survival probabilities--- particularly due to the complexity of random-walk behavior--- is non-trivial. Instead, this study fits for scattering lengths by interpolating between precomputed simulation sets at different regions in the scattering length parameter space.

\item PMT Timing Response--- Similarly to the methods used to fit for the scintillation pulse shape, the timing distribution of PMT pulses and post-pulses can be modified analytically. We utilize the Gumbel distribution~\cite{gumbel1958statistics} used to characterize the main and secondary PMT pulses. Photons are re-weighted analytically by the ratio of the sum of Gumbel distributions describing the PMT response as a function of time, $P(t, \mu, \sigma) = \frac{1}{\sigma} u(t) e^{-u(t)}$ for $u(t) = e^{-(t - \mu) / \sigma}$ and the location of the pulse is $\mu$ and shape is $\sigma$.
\end{enumerate}

In this fit, data and simulation are binned in time, separately for each PMT, between $-30~\text{ns}$ and $2~\mu\text{s}$. From $-30~\text{ns}$ to $-10~\text{ns}$, a $5~\text{ns}$ grid is used for binning, from $-10~\text{ns}$ to $80~\text{ns}$, a $2~\text{ns}$ grid is used to capture fine grained time structure of the rising and falling edge of the pulse shape, and the from $80~\text{ns}$ to $2~\mu\text{s}$ a $20~\text{ns}$ grid binning was utilized to account for lower statistics in the slowly varying triplet region. Information from 191 of 200 PMTs is included, with the 9 remaining PMTs excluded because of elevated noise rates, unstable performance, or other persistent issues. The likelihood comparison uses an effective likelihood treatment designed for limited Monte Carlo sample sizes~\cite{Arguelles:2019izp}, and the parameters presented in subsequent sections are obtained by minimizing the negative log-likelihood with the gradient descent optimization algorithm, \texttt{L-BFGS-B}~\cite{minimizer1,minimizer2}.

\subsection{\label{sec:fit_uncertainties}Systematic Uncertainty Estimation}
Typically in this type of analysis, confidence intervals would be constructed from a profile likelihood ratio test-statistic and approximated according to Wilks' theorem~\cite{Wilks:1938dza}. However, lingering disagreements between data and simulation can cause this method of uncertainty estimation to produce un-physically small errors on parameters. Most of the disagreement between data and simulation manifests as spatially correlated normalization differences across PMTs. In light of this, we adopt an error estimation procedure that examines the level of change required from the fit parameters to resolve the aforementioned discrepancies. Specifically, we run the optical model fit separately for each of the 191 PMTs to obtain a distribution of the best-fit model parameters. We then assign $1\sigma$ uncertainties to the parameters based on the marginal highest posterior density (HPD) region for each parameter. Uncertainty bands for the simulation prediction are similarly computed from the HPD of expectation values resulting from the 191 best-fit points.

\section{\label{sec:fit_results} Fit Components and Results}

This section describes the resulting best fit expectation and elements of the fit in detail, providing values and uncertainties for the model parameters used. We also identify parameters that were set using external information.

\subsection{Overall Time Structure}

\begin{figure}[h]
  \centering
  \includegraphics[width=\linewidth]{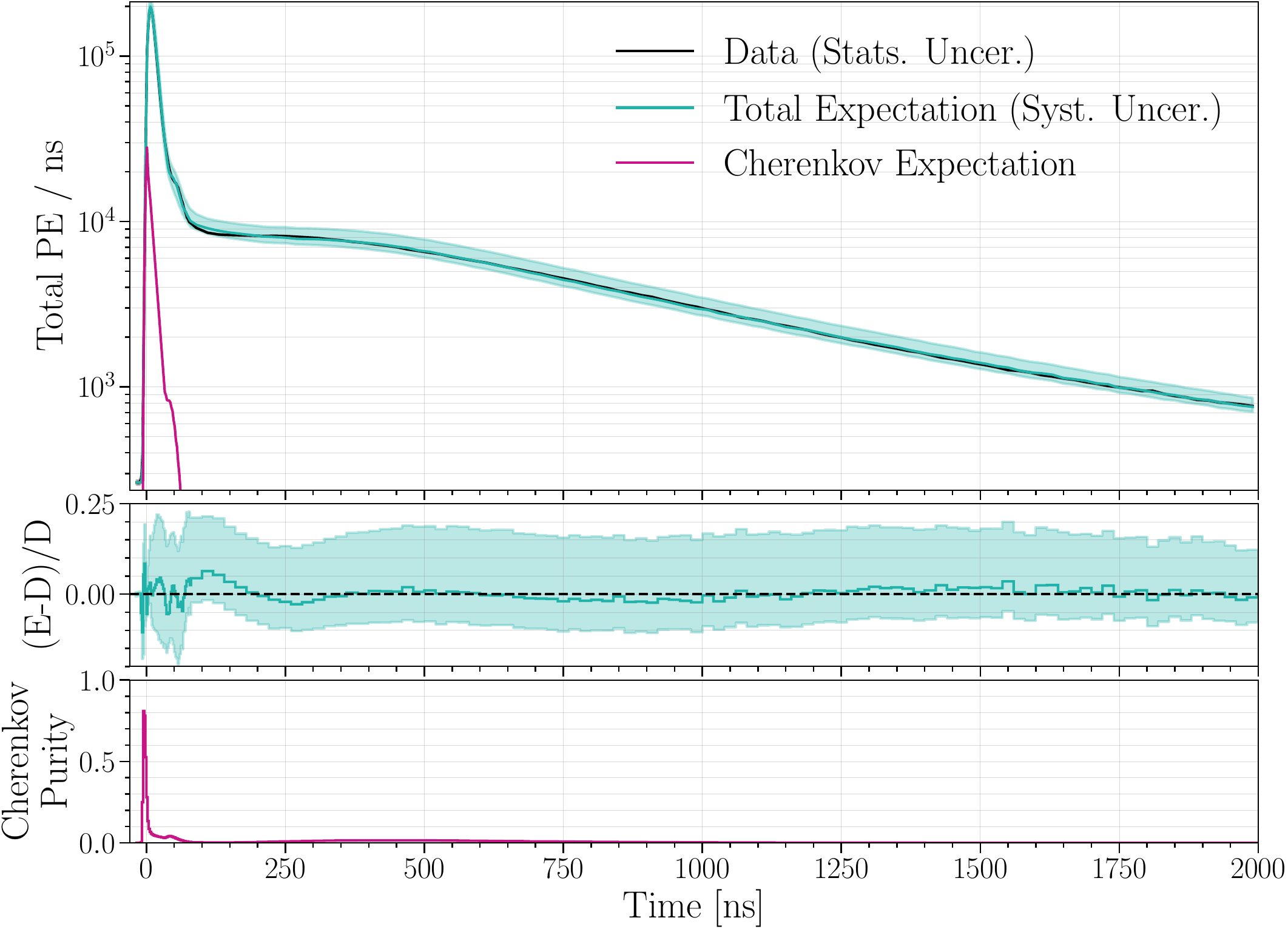}
  \caption{Data (black line) and total expectation (blue line) summed across all PMTs. The Cherenkov contribution to the expectation is demonstrated by the magenta line. The data, with statistical error bands, and the total expectation, with systematic error bands, agree $\leq$10\%. At longer time scales, 250~ns to 2~$\mu$s, the agreement between the data and central value of the expectation is $\leq$5\%. Additionally, at these longer time scales, one of the PMT post-pulse distributions can be seen around 400~ns to 500~ns, see Sec.~\ref{sec:pmt_pulsing_fit} for more discussion.}
  \label{fig:long time scale}
\end{figure}

Although the bulk of the light produced is by scintillation, this work must include the Cherenkov contribution in order to accurately describe the time structure. Fig.~\ref{fig:long time scale}, top, shows the data and the best fit expectation summed across all PMTs across the entire time region used for fitting. This plot also shows the Cherenkov radiation contribution to the total expectation separately. The middle plot presents the residual between data and the total expectation to show the percent deviations as a function of time. Across the entire time region, the data and the predicted values agree within $10\%$. The bottom plot shows the ratio of expected Cherenkov photons to total expected photons, referred to as the ``Cherenkov purity.'' 

\begin{figure}[h]
  \centering
  \includegraphics[width=\linewidth]{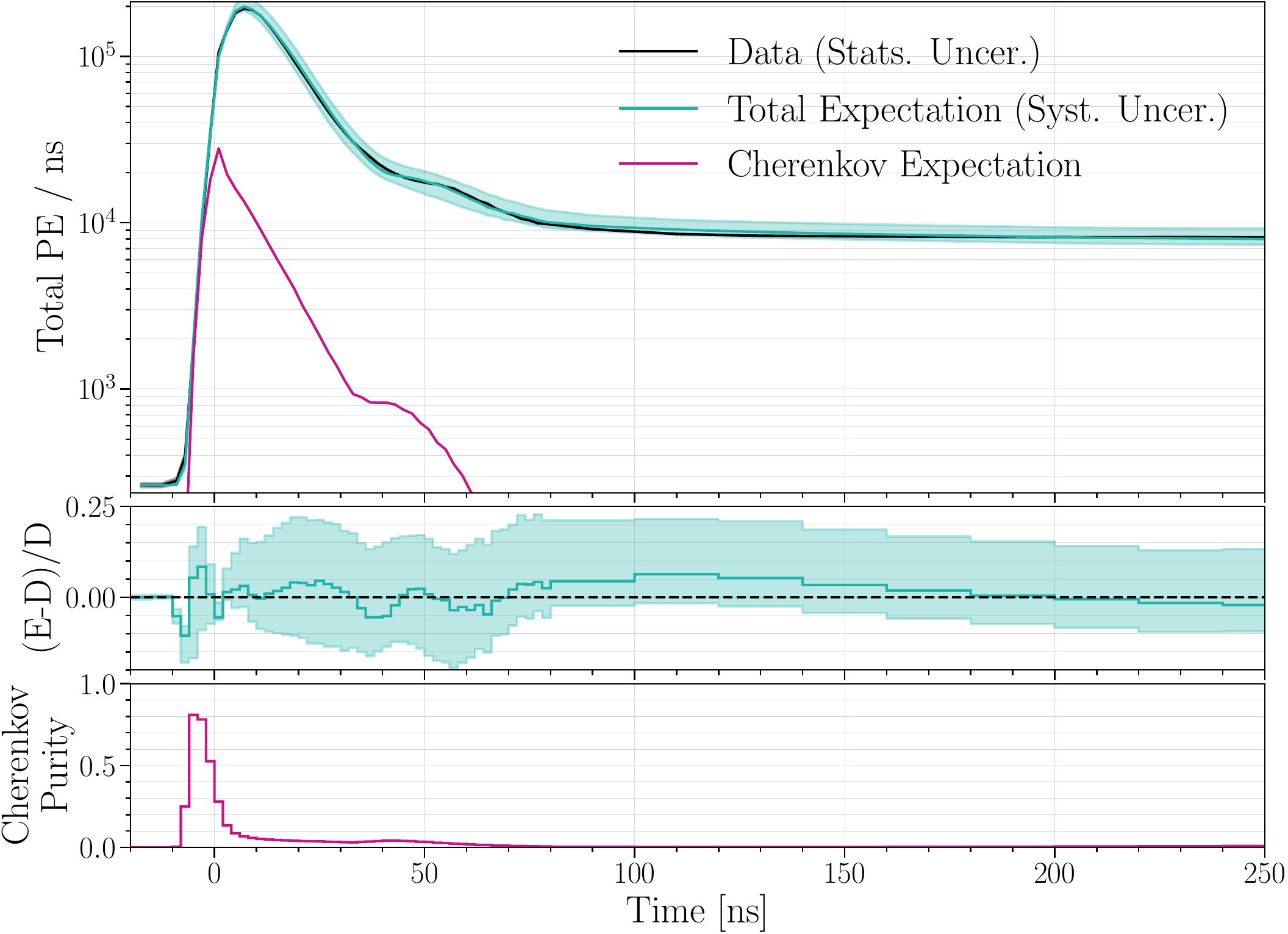}
  \caption{Zoomed in view of Fig.~\ref{fig:long time scale}. This shows the peak of the accumulated PE pulse series in better detail. There is also evidence of an artifact of the PMT post-pulsing around 50~ns. We are able to characterize this region of the expectation to agree with data within $\leq$5\%.}
  \label{fig:medium time scale}
\end{figure}

\begin{figure}[h]
  \centering
  \includegraphics[width=\linewidth]{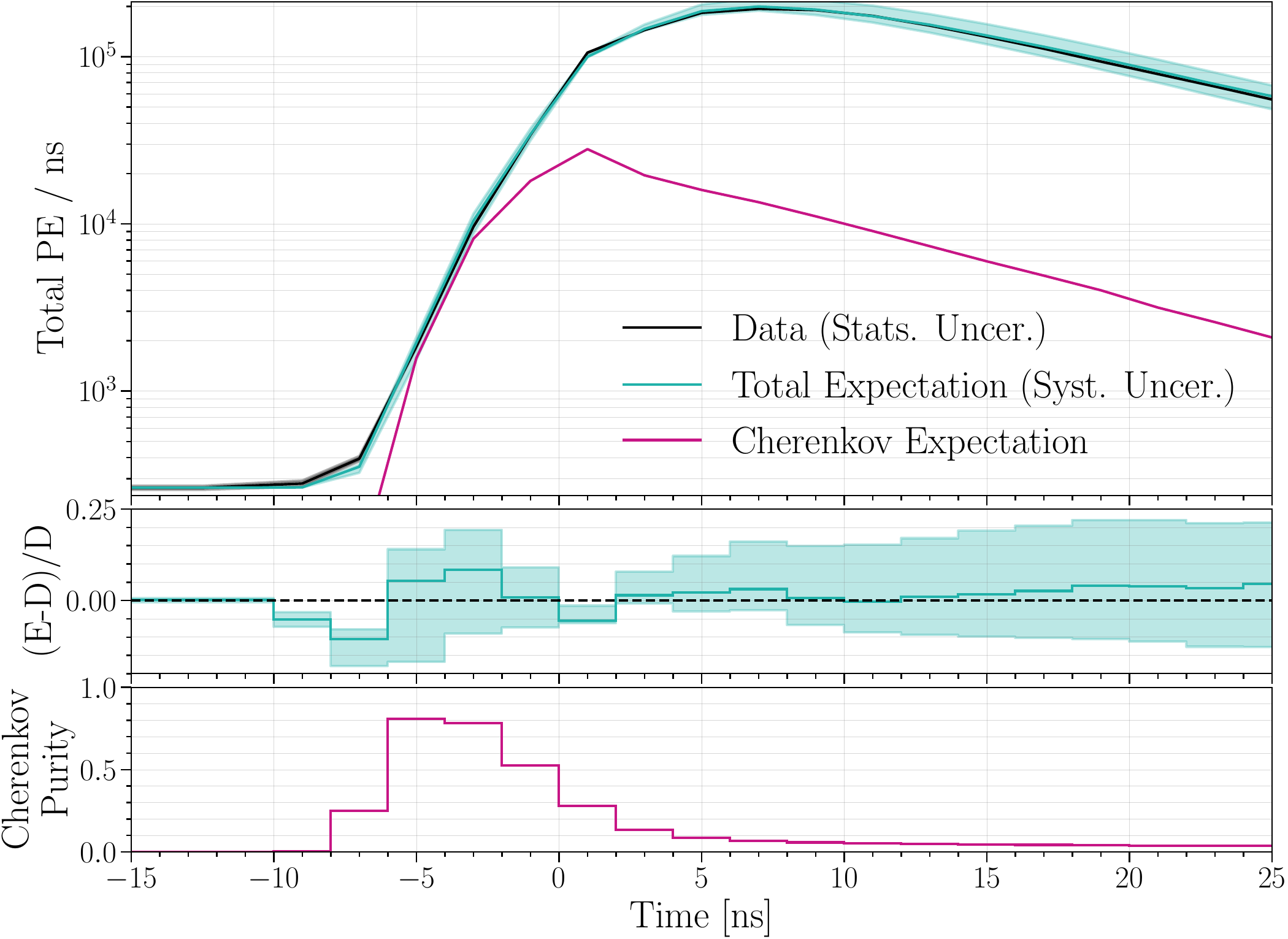}
  \caption{Final view of Fig.~\ref{fig:long time scale} with emphasis on the PE pulse series close to the event start time at $t=0$. For $-6 \leq t < 0$~ns, the Cherenkov purity is $>50\%$ in the entire detector. The Cherenkov purity does not drop below 10\% until $t > 5$~ns, necessitating full description of both scintillation and Cherenkov light production and propagation to accurately characterize the detector response. For more details of the Cherenkov expectation, see Ref.~\cite{companion_paper}.}
  \label{fig:short time scale}
\end{figure}

For the longer-time scale structure, 250~ns to 2~$\mu$s, there is better than 5\% agreement between data and the central value of the expectation. For the medium time scale agreement, Fig.~\ref{fig:medium time scale} shows the data and expectation in a shorter time window. PMT post-pulsing is evident around 50~ns. Data and expectation agree within $\leq 5\%$ around 50~ns, indicating that this PMT post-pulsing model is reliable. For the short time scale agreement, Fig.~\ref{fig:short time scale} shows the data and expectation very close to the reconstructed event start times. This time region is where the Cherenkov purity is most evident. Across all PMTs, Cherenkov light is a $>10\%$ component of the total expectation until $t\approx 5~\rm{ns}$ after the event start time. This necessitates simultaneous characterization of both scintillation and Cherenkov photons to fully understand light production and propagation in LAr. For a full discussion of the expectation in the early time region, especially in terms of the Cherenkov light identification, see Ref.~\cite{companion_paper}.

\subsection{Scintillation Light Parameters}

\begin{table}[b]
  \caption{Scintillation pulse shape fit parameters with uncertainties. See Eq.~\ref{eq:lar_pulse_shape} for full description of the scintillation time dependence.}
  \begin{ruledtabular}
    \begin{tabular}{ccc}
      Parameter & Central Value &  Uncertainties \\
      \hline
      $R_s$ & $0.367$ & $-0.015, +0.017$ \\
      $R_t$ & $0.633$ & $-0.015, +0.017$ \\
      $\tau_s$ & $4.28~\mathrm{ns}$ & $-0.42~\mathrm{ns}, +1.20~\mathrm{ns}$ \\
      $\tau_t$ & $588.80~\mathrm{ns}$ & $-3.30~\mathrm{ns}, +3.65~\mathrm{ns}$ \\
    \end{tabular}
  \end{ruledtabular}
  \label{table:light_table} 
\end{table}

Table~\ref{table:light_table} describes the best fit central value and uncertainties for the scintillation light production parameters. We explored allowing an intermediate time component, but the fit across all PMTs preferred only the singlet and triplet time components. For this reason, we fixed the total ratio of singlet and triplet light to unity for uncertainty estimation. From the optimization, we find preference for $0.367^{+0.017}_{-0.015}$ ratio of singlet light, $0.633^{+0.017}_{-0.015}$ ratio of triplet light,
$4.28^{+1.20}_{-0.42}$~ns singlet time constant, and $588.80^{+3.65}_{-3.30}$~ns triplet time constant. The effects of the impurities in the LAr are most evident through the reduction of the triplet time constant~\cite{WArP:2008dyo,WArP:2008rgv,Jones:2013bca}. Quenching of triplet light also increases the expected intensity of singlet light component. We can characterize both of these effects, as we would expect for ppm-level oxygen, nitrogen, and water contamination in LAr. 

\subsection{\label{sec:tpb}TPB Characteristics}
The TPB is fully modeled and described in~\texttt{Geant4} using measurements of the re-emission spectrum and absorption length~\cite{Benson:2017vbw,doi:10.1021/j100052a011}. We investigated the impacts of deviations to the absorption length, especially in the exponential tail that overlaps with the re-emission spectrum. We found that this did not significantly impact the fit, so we fixed the absorption length and re-emission spectrum to the literature values.

We also investigated the effects of a TPB wavelength-shift time constant. Through simulation of short time-scale wavelength-shift time constants, we found that an exponential time constant of 0.3~ns reflected the data most accurately. To simplify subsequent fits for other light parameters, we fixed the TPB time constant to that value. While there is evidence for longer-scale time structure in the TPB re-emission, since we are fitting to a relatively short time region of 2~$\mu$s after the event start time, we neglected this effect~\cite{DEAP:2020hms}. 

\subsection{\label{sec:uv_abs}Photon Absorption}
While pure LAr is transparent to photons with wavelengths above the first excimer continuum of approximately 113~nm, impurities can quench the scintillation process and absorb scintillation photons~\cite{Neumeier:2012cz}. This absorption is highly dependent on the amount and types of impurities in the LAr, resulting in different measurements of absorption lengths from different experiments. Measurements of absorption lengths integrated over UV wavelengths vary from 66~cm to 50~cm~\cite{ISHIDA1997380, ArDM:2016jbw}. Other investigations reported that transmission through purified LAr depended strongly on wavelength, necessitating a wavelength resolved function for the absorption length~\cite{Neumeier:2012cz,Neumeier:2015lka,Fields:2020wge}. 

\begin{figure}[h]
  \centering
  \includegraphics[width=\linewidth]{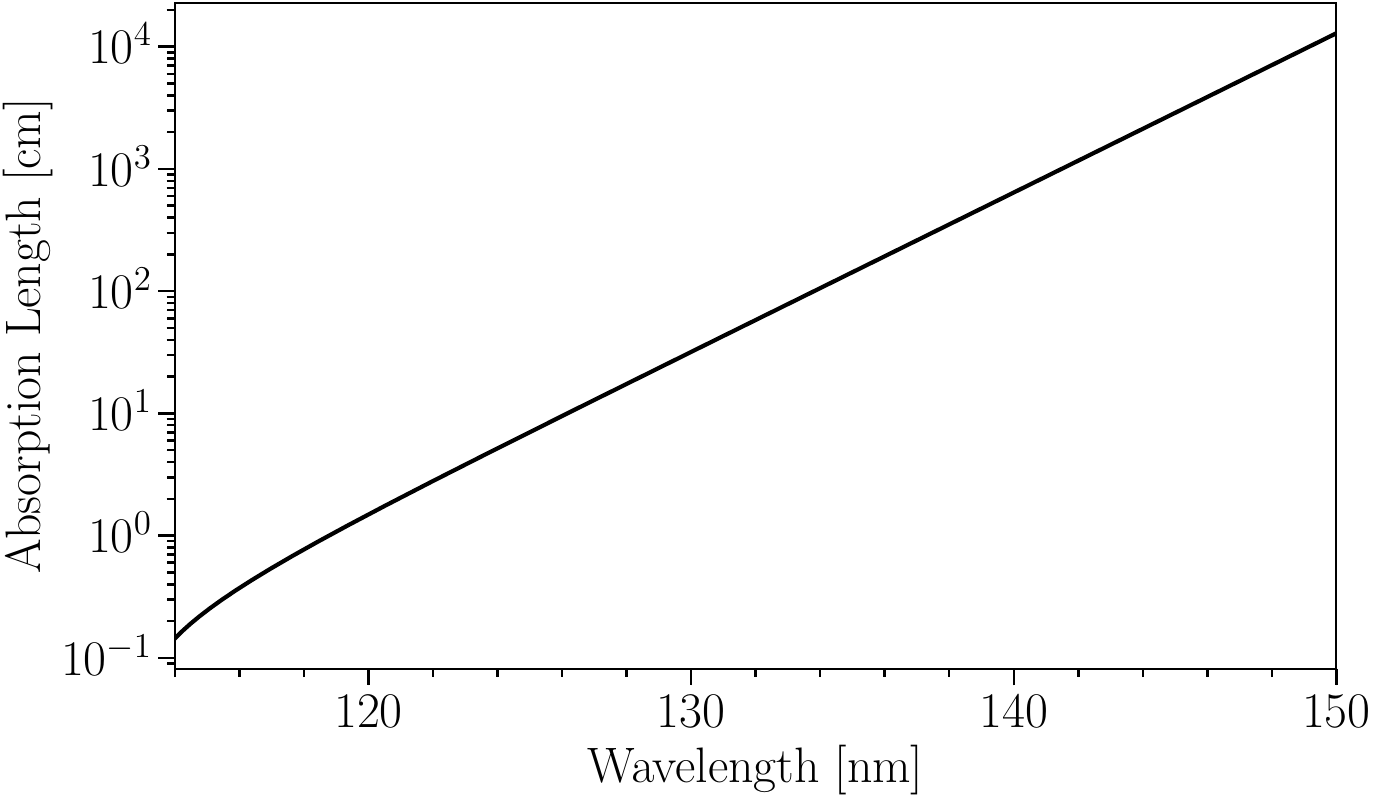}
  \caption{Preferred absorption length of photons in LAr between 113~nm and 150~nm. This absorption length, see Eq.~\ref{eq:uv_absorption}, is parametrized by Ref.~\cite{Neumeier:2012cz} to allow for wavelength resolved absorption. At 128~nm, the absorption length is 17.42~cm. We do not report uncertainties on this parameter because the method for estimating uncertainties, described in Sec.~\ref{sec:fit_uncertainties}, removes distance-dependent information and this parameter becomes degenerate with normalization.}
  \label{fig:uv_abs}
\end{figure}

Using the formalism in Ref.~\cite{Neumeier:2012cz}, Eq.~\ref{eq:uv_absorption} describes the parametrization of wavelength dependent absorption length we used in optimizing the optical model. We allowed the parameters $a$ and $d$ to vary, to describe changes to the shape and normalization, and fixed parameter $b$ to the literature value of 113.2~nm. Fig.~\ref{fig:uv_abs} demonstrates the optimized wavelength resolved absorption length. From the best fit parameters of $a = 0.30$ and $d = 0.194$, at 128~nm we find an absorption length of 17.42~cm which then increases exponentially. Additionally, we allowed for absorption in the near-UV visible wavelengths due to impurities. We found preference for 98.25~cm absorption length integrated from 300~nm to 400~nm.

\begin{equation}
\lambda_{att} = \frac{d}{\ln \left(\frac{1}{1 - e^{-a (\lambda - b)}} \right)}
\label{eq:uv_absorption}
\end{equation}

While our uncertainty estimation is useful for understanding the deviations in the time structure of the expectation across all the PMTs, it does not quantify parameters that effectively control the normalization of the expectation. Without the distance dependent constraint of fitting to all PMTs simultaneously, uncertainties on the photon absorption length from individual fits are artificially very large and so we do not quote them. 

\subsection{\label{sec:birk}Birks' Law}
Scintillation light yield is not proportional to the energy deposited per unit distance due to a non-linear quenching factor from the exciton formation~\cite{Birks:1951boa}. This can be described by Birks' Law, Eq. \ref{eq:birk}, where $dL / dx$ is the scintillation light yield per unit distance,  $S$ is the scintillation efficiency, $dE / dx$ is the energy deposited per unit distance, and $k$ is Birks' coefficient~\cite{Birks:1951boa}.

\begin{equation}
\frac{dL}{dx} = \frac{S \cdot \frac{dE}{dx}}{1 + k \cdot \frac{dE}{dx}}
\label{eq:birk}
\end{equation}

This has been experimentally found to agree with scintillation in LAr~\cite{AMORUSO2004275}. The ICARUS collaboration measured Birks' constant $k = 0.0486 \pm 0.0006$~(kV/cm)((g/cm$^2$)/MeV) across electric drift fields $0.1 < E < 1.0$~kV/cm using a modification of Birks' law for the presence of an electric field $\varepsilon$, detailed in Eq.~\ref{eq:e_field_birk} where $Q$ and $Q_0$ denote the initial ionization and collected charges, respectively. 

\begin{equation}
Q = \frac{Q_0}{1 + \frac{k}{\varepsilon} \cdot \frac{dE}{dx}}
\label{eq:e_field_birk}
\end{equation}

This relationship breaks down as $\varepsilon \rightarrow 0$, making ICARUS's measurement of Birks' constant inapplicable in a detector like CCM without any electric drift field. Instead, we used the linear energy transfer (LET) measurement for 1~MeV electrons in LAr with no electric field to calculate Birks' constant, detailed in Eq.~\ref{eq:scint_eff_birk}~\cite{Doke:1990rza}. 

\begin{equation}
LET = \frac{1}{1 + k \cdot \frac{dE}{dx}}
\label{eq:scint_eff_birk}
\end{equation}

For LAr at 87$^\circ$K, the stopping power of 1~MeV electrons is 1.386~MeV cm$^2$/g~\cite{PDG2024}. This leads to a Birks' constant for 1~MeV electrons in LAr without an electric field of $k = 0.295$~(g/cm$^2$)/MeV. We studied the effects of altering Birks' constant and found that it was degenerate with overall scintillation normalization for these low energy sodium decay events and therefore we fix its value. Future calibration studies of higher energy Michel electrons from muon decay in CCM200 may be able to directly probe Birks' constant in LAr detectors without an applied electric field.

\subsection{\label{sec:rindex}Index of Refraction}
\begin{figure}[h]
  \centering
  \includegraphics[width=\linewidth]{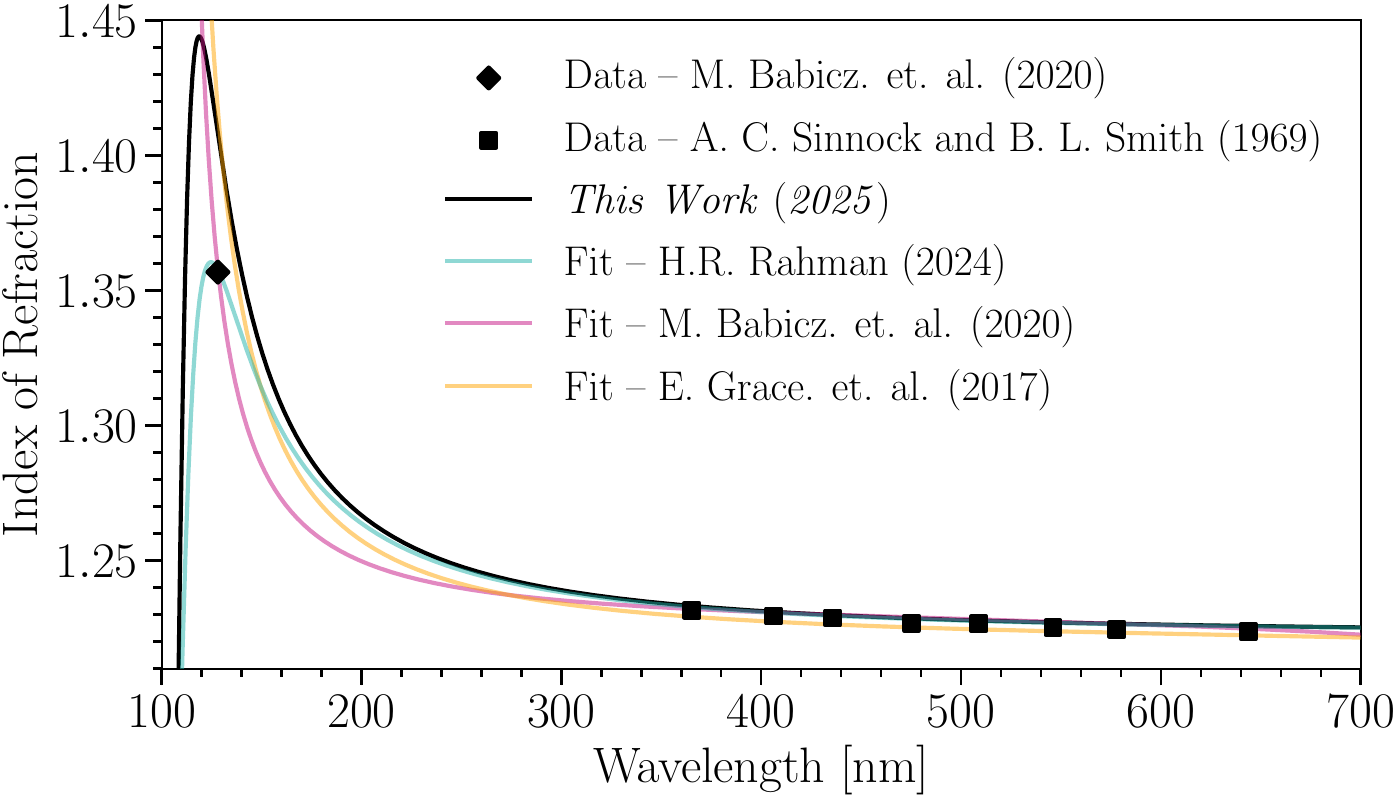}
  \caption{Comparison of literature values of the index of refraction in LAr with the results of this work. The data points and previous fits, described in Sec.~\ref{sec:lar_cherenkov_physics}, are plotted for comparison with the results of this work (black line). This fit prefers a smoother transition from the index of refraction in the VUV region to the visible region. Without any data between between 128~nm and 350~nm, this region cannot be well constrained.}
  \label{fig:best_fit_rindex}
\end{figure}

As described in Sec.~\ref{sec:lar_cherenkov_physics}, the index of refraction is an important parameter in Cherenkov radiation production as well as photon interaction in a medium. While the index of refraction of visible photons in LAr is well measured between 350~nm and 650~nm~\cite{Sinnock:1969zz}, there is only one experimental constraint in the VUV range~\cite{Babicz:2020den}. We allowed the $\gamma_{UV}$ parameter in the damped harmonic oscillator fit for index of refraction, described in Eq.~\ref{eq:ho_rindx}, to vary, as it primarily affects the VUV-UV region of parameter space. We found a preferred value of $\gamma_{UV}=0.0018$, see Fig.~\ref{fig:best_fit_rindex} for the best fit index of refraction. For simplification in the fitting process, we fix the index of refraction to this parameterization while exploring other parameters and do not report an uncertainty on this parameter. Future measurements of the index of refraction at wavelengths $<350~\rm{nm}$, especially around 200~nm, would be very helpful in constraining the behavior between the VUV regime and the visible regime. 

\subsection{\label{sec:rayl}Rayleigh Scattering}
\begin{figure}[h]
  \centering
  \includegraphics[width=\linewidth]{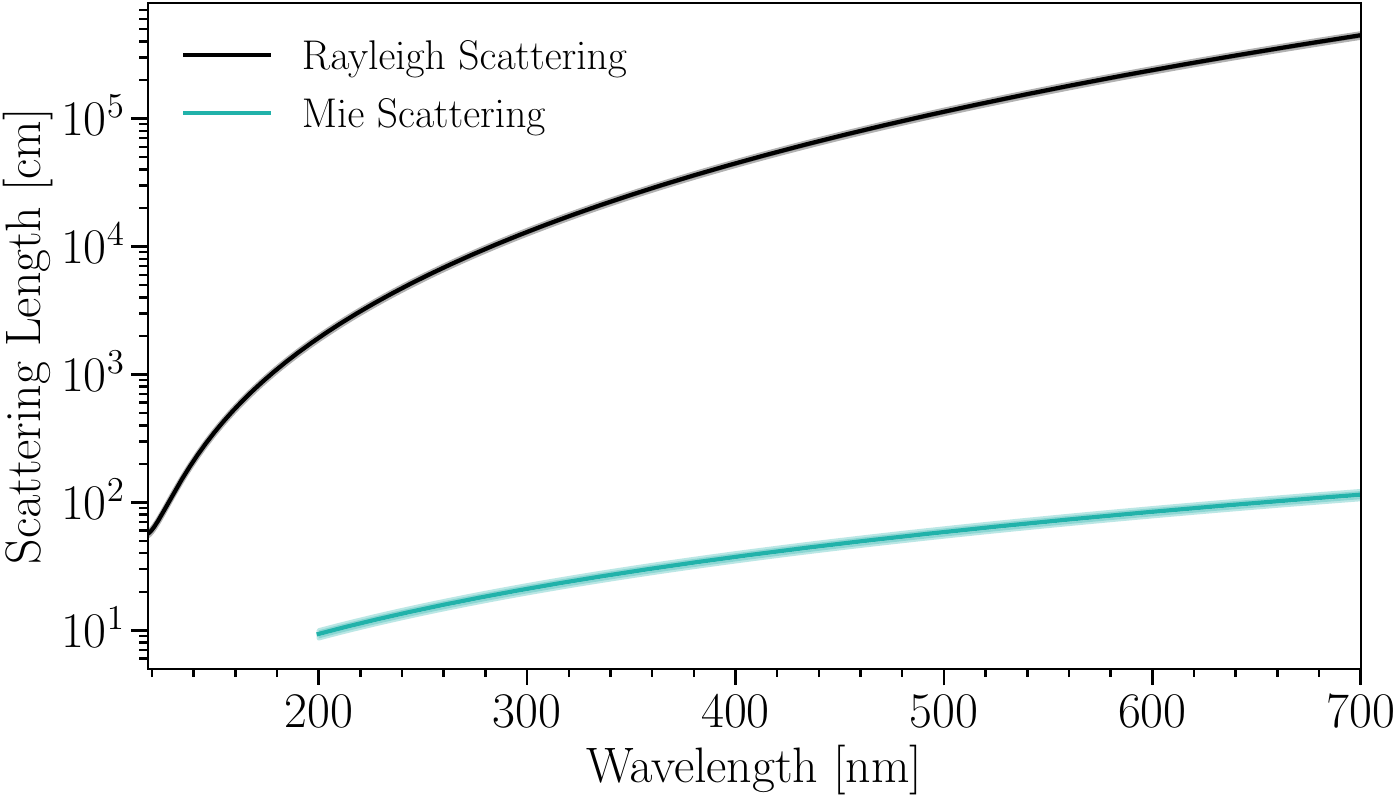}
  \caption{Rayleigh scattering length (black) and Mie scattering length (blue) with systematic uncertainties. At 128~nm, this fit prefers a Rayleigh scattering length of $99.98^{+3.56}_{-4.52}~\rm{cm}$. The Rayleigh scattering has the expected dependence on the index of refraction and goes as $1 / \lambda^4$, described in Eq.~\ref{eq:rayl}. The Mie scattering length, allowed for wavelengths above 200~nm due to the requirement that the wavelength is comparable to the size of the scattering particle, is $9.37^{+0.63}_{-0.73}~\rm{cm}$ at 200~nm. This is the first characterization of Mie scattering in LAr, motivations described in Sec.~\ref{sec:mie}. Measurements resolving the Mie scattering cross section for different possible impurities in LAr would be useful to improve this characterization.}
  \label{fig:rayl_mie_scattering}
\end{figure}

Argon atoms inherently Rayleigh scatter optical photons due to their size being much smaller than the photon wavelengths. The index of refraction is related to the Rayleigh scattering length, described in Eq. \ref{eq:rayl}~\cite{landau1984electrodynamics}. Calculations and measurements of Rayleigh scattering length in LAr at 128~nm vary between 55~cm and 100~cm~\cite{Grace:2015yta,Babicz:2020den,Seidel:2001vf}. We fit for a Rayleigh scattering length of $99.98^{+3.56}_{-4.52}~\rm{cm}$ at 128~nm, shown in Fig.~\ref{fig:rayl_mie_scattering}. The Rayleigh scattering length may be effectively longer in the CCM detector due to contamination increasing the probability of absorption for short-wavelength photons that travel further distances before wavelength shifting. 

\begin{equation}
l^{-1} = \frac{16 \pi^3}{6 \lambda^4} \left[ k T \rho^2 \kappa_T \left(\frac{(n^2-1)(n^2+2)}{3} \right)^2 \right]
\label{eq:rayl}
\end{equation}

\subsection{\label{sec:mie}Mie Scattering}
While Rayleigh scattering is a strong effect for VUV photons, due to the $1 / \lambda^4$ dependence, Mie scattering predominantly affects photons with larger wavelengths and has a distinct forward angular dependence~\cite{Mie1908}. When the particle size approaches the wavelength of the incoming optical photon, Mie scattering becomes a relevant process. While detectors with pure LAr would not expect Mie scattering given the small size of the argon atoms, impurities could be a source of Mie scattering targets in CCM, particularly due to the lack of filtering.

There are several candidates for Mie scattering centers in the CCM200 detector. First, TPB evaporative coating on foils and PMTs has been demonstrated to produce particulates in LAr~\cite{Asaadi:2018ixs,Ignarra:2014yqa}. The large crystal structure of TPB would facilitate optical photons Mie scattering in LAr. Second, since CCM200 was not constructed in a dedicated clean room, though diligence was taken to keep all detector elements clean, some dust may remain from the construction process. Third, more speculatively, there is the known water impurity of 0.01~ppm as quoted by the manufacturer that could form ice crystals. Lastly, similarly speculatively, as the LAr boils off, it could create bubbles in the fiducial region that can cause Mie scattering.   

Because the exact content of each of these sources in the LAr, as well as potential contributions from other unknown sources, is very difficult to measure in CCM, we fit for an average Mie scattering effect. Since the cross section for Mie scattering depends on the target size and index of refraction, and there many possible sources of Mie scattering in CCM, we used $1 / \lambda^2$ scaling dependence to fit for an effective Mie scattering length affecting photons with wavelengths above 200~nm. Fig.~\ref{fig:rayl_mie_scattering} demonstrates the preferred Mie scattering length with uncertainties, which is $9.37^{+0.63}_{-0.73}~\rm{cm}$ at 200~nm. Further study of Mie scattering in LAr, especially its impact on directional reconstruction, could improve the physics modeling of optical photons.

\subsection{\label{sec:pmt_pulsing_fit}PMT Timing Characterization}
The final effect that was considered to impact the time distribution of expected PEs is the intrinsic PMT time distribution. Post-pulsing occurs for many models of PMTs~\cite{MiniBooNE:2006fhd,Kaptanoglu:2017jxo,DEAP:2017fgw,Abbasi_2010,Brigatti_2005}, including CCM's. Some sources of post-pulsing arise when PEs back scatter off of one of the dynode stages, introducing a time delayed digitized signal. Longer time-scale post-pulsing effects can be caused by ionization of heavy elements in the PMT that then travel through the same amplification chain, with a much slower velocity than electrons, and this leads to digitized signals at long time scales after the main PMT pulse.

\begin{figure}[h]
  \centering
  \includegraphics[width=\linewidth]{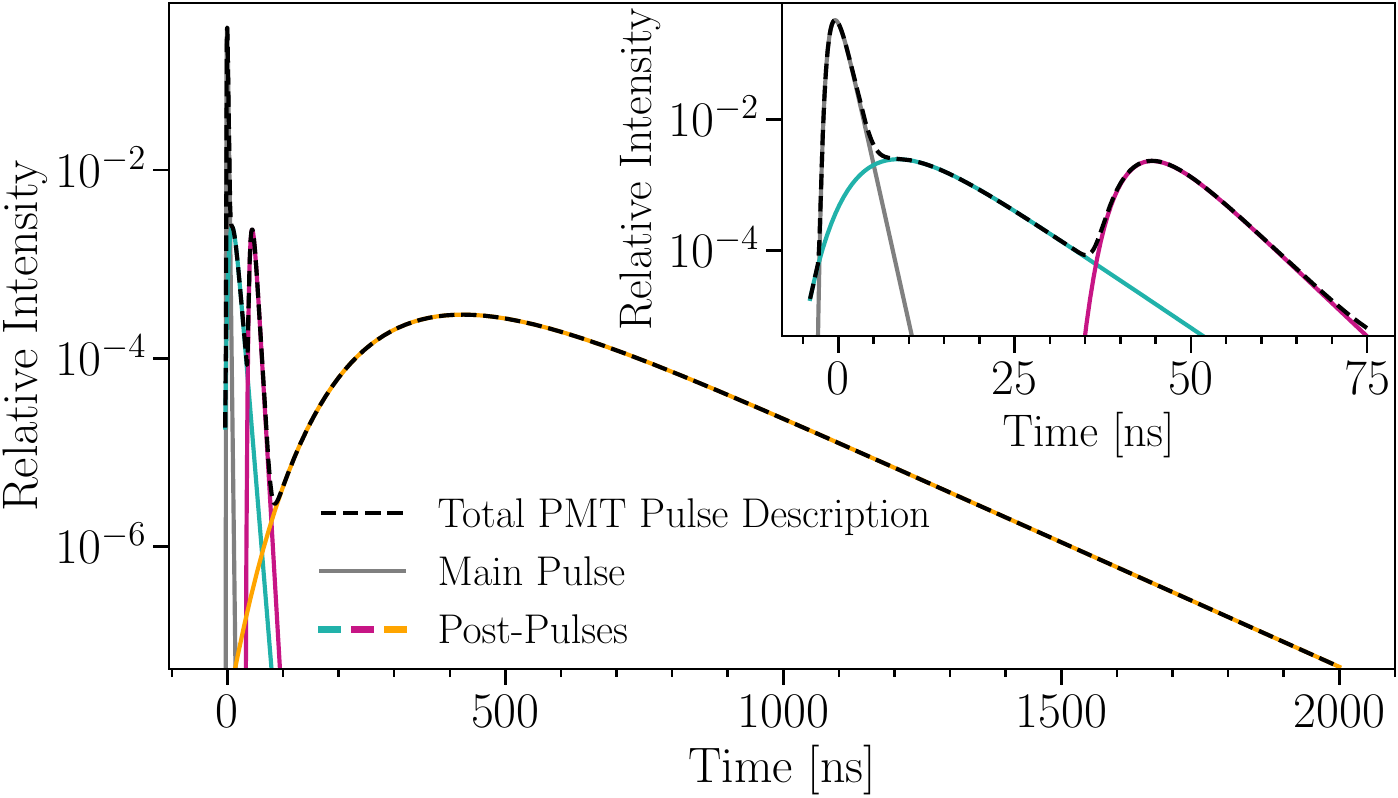}
  \caption{PMT timing characterization. All of the PMT pulses are described using a Gumbel distribution~\cite{gumbel1958statistics}. The main PMT pulse, centered at -0.4~ns and with a width of 0.9~ns, describes the transit time spread of the R5912-Y002 PMTs. This was fixed in the characterization of the PMT timing response. The three additional PMT post-pulses can be seen in the blue, pink, and orange lines. The blue and pink PMT post-pulses are centered before 50~ns, see inset plot. The final PMT post-pulse is centered at 423.24~ns and modifies the triplet expectation. See Table~\ref{table:pmt_params} for full description of the PMT post-pulse parameters.}
  \label{fig:best_fit_late_pulse}
\end{figure}

Because of the deviations to the expected time structure, this work requires characterization of the PMT pulses in addition to LAr photon creation and propagation physics. We use a Gumbel distribution to describe the expected timing of the PMT pulses. We found preference for three PMT post-pulses, plotted in Fig.~\ref{fig:best_fit_late_pulse} and parameters described in Table~\ref{table:pmt_params}, in addition to the fixed timing description for the main PMT pulse based on the PMT electron transit time. The inset plot shows a zoomed in view of the time structure expected in the first 75~ns of a PMT pulse. There are two preferred PMT post-pulses in this region, one around 8.47~ns and another at 44.51~ns. Additionally, there is preference for a PMT post-pulse at 423.24~ns, which can be seen to affect the shape of the triplet expectation. We find that post-pulsing is a 19.7\% effect compared to the main PMT peak and note that the PMTs in CCM200 are typically operated at 1700~V for desired gain and have a typical current draw of 750~$\mu$A. 

\begin{table*}
  \caption{PMT timing response fit parameters and uncertainties. We allowed up to four PMT post-pulses but found preference for only three, plotted in Fig.~\ref{fig:best_fit_late_pulse} along with the main PMT pulse to show the total PMT timing distribution. Main PMT pulse parameters are derived from the spread in electron transit times for R5912-Y002 PMTs and were fixed in this work.}
  \begin{ruledtabular}
    \begin{tabular}{cccc}
      Pulse & Location~[ns] &  Shape~[ns] & Probability \\
      \hline
      Main Pulse & $-0.45$ & $0.9$ & $0.803$ \\
      Post-Pulse 1 & $8.47~[+2.42, -1.20]$ & $6.00~[+2.12, -1.52]$ & $0.040~[+0.038, -0.040]$ \\
      Post-Pulse 2 & $44.51~[+1.00, -7.21]$ & $4.26~[+4.77, -0.21]$ & $0.027~[+0.017, -0.008]$ \\
      Post-Pulse 3 & $423.24~[+0.63, -1.22]$ & $163.84~[+1.71, -1.49]$ & $0.13~[+0.028, -0.001]$ \\
    \end{tabular}
  \end{ruledtabular}
  \label{table:pmt_params} 
\end{table*}

\section{\label{sec:applications}Applications of this Model}

The immediate application for the model described in this paper is for physics analyses on the CCM experiment. The CCM physics program consists of low-energy BSM searches using coherent scattering signal events at the $\mathcal{O}(100)$~keV scale, as well as a higher energy BSM searches and cross section measurements using electromagnetic signal final state events in the $\mathcal{O}(10)$~MeV to $\mathcal{O}(100)$~MeV energy range. This model has been tuned for the low energy program, and will be leveraged to match Michel electron data for the high energy program in a future study. 

This work is the first characterization of LAr scintillation in a large light collection only detector to ascribe uncertainties on the scintillation light parameters. As such, this paper provides a template for future studies by other detectors. This model can also be used to explore the design of larger detectors based on the CCM200 detector. In particular, the need-versus-cost for relatively expensive purification systems can be explored using this model. We note that the relatively short lifetime of the triplet state in the presence of impurities is an advantage in that it leads to fewer single PE backgrounds from previous events. This is especially important for applications in high-rate environments such as spallation neutron sources, where reconstruction of physics events while there is a high rate background (from steady state cosmogenic muons, ambient radioactivity, and low energy neutrons) can be strongly affected by residual triplet photon backgrounds. The impurities also absorb the UV photons preferentially, reducing the total early scintillation light and thereby allowing a larger Cherenkov-to-scintillation ratio. 

This work forms the basis for accurately calibrating a Monte Carlo simulation used with large light collection detectors, which is essential for Cherenkov light separation. As demonstrated in Ref.~\cite{companion_paper}, Cherenkov light identification relies on detailed understanding of the time structure of PE arrivals. This work creates a template for investigating the effects of a large number of optical parameters in such detectors, with the goals of event-by-event Cherenkov light separation and the creation of a high-fidelity detector response model using differentiable simulation.  

\section{\label{sec:conclusion}Conclusion}

LAr is a common target and detection material for neutrino and weakly interacting physics searches. While many LAr detectors utilize the charge readout of TPCs for $>$100~MeV event reconstruction, CCM is the largest detector by mass to collect only the photons emitted in the LAr for keV to MeV scale event reconstruction. This requires a well tuned Monte Carlo simulation of both light production and propagation. We explore parameters related to scintillation light emission, index of refraction, photon absorption, TPB absorption and re-emission, Birks' law photon quenching, Rayleigh scattering, Mie scattering, and PMT time structure in order to tune an optical model. Since there are more than 20 parameters that went into this work, we utilized differentiable simulation to evaluate the expectation as a function of the physics parameters, allowing for efficient generation of the expectation and gradient-based optimization. After optimization of the optical model parameters, we obtained deviations between data and the central value of the expectation of $\leq10\%$. Additionally, we ascribed uncertainties through the variation of expectation across the PMTs in the detector. This work motivates further study of Mie scattering in LAr and fuller understanding of the effects of impurities. Using this simulation as a foundation, this work enabled identification of Cherenkov light on an event-by-event basis produced by sub-MeV electrons, fully described in Ref.~\cite{companion_paper}.

\begin{acknowledgments}
We acknowledge the support of the Los Alamos National Laboratory LDRD and the U.S. Department of Energy Office of Science funding. We also wish to acknowledge support from the LANSCE Lujan Center and LANL’s Accelerator Operations and Technology (AOT) division. This research used resources provided by the Los Alamos National Laboratory Institutional Computing Program, which is supported by the U.S. Department of Energy National Nuclear Security Administration under Contract No.~89233218CNA000001. DAN is supported by the NSF Graduate Research Fellowship under Grant No.~2141064. AAA-A, CFMA, JCD, and MCE acknowledge support from DGAPA-UNAM Grant No.~PAPIIT-IN104723. We thank Vassili Papavassiliou, New Mexico State University, for input on modeling the index of refraction in the VUV range.
\end{acknowledgments}

\bibliographystyle{apsrev4-2}
\bibliography{main}

\begin{thebibliography}{78}%
\makeatletter
\providecommand \@ifxundefined [1]{%
 \@ifx{#1\undefined}
}%
\providecommand \@ifnum [1]{%
 \ifnum #1\expandafter \@firstoftwo
 \else \expandafter \@secondoftwo
 \fi
}%
\providecommand \@ifx [1]{%
 \ifx #1\expandafter \@firstoftwo
 \else \expandafter \@secondoftwo
 \fi
}%
\providecommand \natexlab [1]{#1}%
\providecommand \enquote  [1]{``#1''}%
\providecommand \bibnamefont  [1]{#1}%
\providecommand \bibfnamefont [1]{#1}%
\providecommand \citenamefont [1]{#1}%
\providecommand \href@noop [0]{\@secondoftwo}%
\providecommand \href [0]{\begingroup \@sanitize@url \@href}%
\providecommand \@href[1]{\@@startlink{#1}\@@href}%
\providecommand \@@href[1]{\endgroup#1\@@endlink}%
\providecommand \@sanitize@url [0]{\catcode `\\12\catcode `\$12\catcode `\&12\catcode `\#12\catcode `\^12\catcode `\_12\catcode `\%12\relax}%
\providecommand \@@startlink[1]{}%
\providecommand \@@endlink[0]{}%
\providecommand \url  [0]{\begingroup\@sanitize@url \@url }%
\providecommand \@url [1]{\endgroup\@href {#1}{\urlprefix }}%
\providecommand \urlprefix  [0]{URL }%
\providecommand \Eprint [0]{\href }%
\providecommand \doibase [0]{https://doi.org/}%
\providecommand \selectlanguage [0]{\@gobble}%
\providecommand \bibinfo  [0]{\@secondoftwo}%
\providecommand \bibfield  [0]{\@secondoftwo}%
\providecommand \translation [1]{[#1]}%
\providecommand \BibitemOpen [0]{}%
\providecommand \bibitemStop [0]{}%
\providecommand \bibitemNoStop [0]{.\EOS\space}%
\providecommand \EOS [0]{\spacefactor3000\relax}%
\providecommand \BibitemShut  [1]{\csname bibitem#1\endcsname}%
\let\auto@bib@innerbib\@empty
%</preamble>
\bibitem [{\citenamefont {Abed~Abud}\ \emph {et~al.}(2024)\citenamefont {Abed~Abud} \emph {et~al.}}]{DUNE:2024qgl}%
  \BibitemOpen
  \bibfield  {author} {\bibinfo {author} {\bibfnamefont {A.}~\bibnamefont {Abed~Abud}} \emph {et~al.} (\bibinfo {collaboration} {DUNE}),\ }\href {https://doi.org/10.1103/PhysRevD.110.092011} {\bibfield  {journal} {\bibinfo  {journal} {Phys. Rev. D}\ }\textbf {\bibinfo {volume} {110}},\ \bibinfo {pages} {092011} (\bibinfo {year} {2024})},\ \Eprint {https://arxiv.org/abs/2408.00582} {arXiv:2408.00582 [hep-ex]} \BibitemShut {NoStop}%
\bibitem [{\citenamefont {Acerbi}\ \emph {et~al.}(2024)\citenamefont {Acerbi} \emph {et~al.}}]{DarkSide-20k:2024yfq}%
  \BibitemOpen
  \bibfield  {author} {\bibinfo {author} {\bibfnamefont {F.}~\bibnamefont {Acerbi}} \emph {et~al.} (\bibinfo {collaboration} {DarkSide-20k}),\ }\href {https://doi.org/10.1038/s42005-024-01896-z} {\bibfield  {journal} {\bibinfo  {journal} {Commun. Phys.}\ }\textbf {\bibinfo {volume} {7}},\ \bibinfo {pages} {422} (\bibinfo {year} {2024})},\ \Eprint {https://arxiv.org/abs/2407.05813} {arXiv:2407.05813 [hep-ex]} \BibitemShut {NoStop}%
\bibitem [{\citenamefont {Abratenko}\ \emph {et~al.}(2024)\citenamefont {Abratenko} \emph {et~al.}}]{SBND:2024vgn}%
  \BibitemOpen
  \bibfield  {author} {\bibinfo {author} {\bibfnamefont {P.}~\bibnamefont {Abratenko}} \emph {et~al.} (\bibinfo {collaboration} {SBND}),\ }\href {https://doi.org/10.1140/epjc/s10052-024-13306-3} {\bibfield  {journal} {\bibinfo  {journal} {Eur. Phys. J. C}\ }\textbf {\bibinfo {volume} {84}},\ \bibinfo {pages} {1046} (\bibinfo {year} {2024})},\ \Eprint {https://arxiv.org/abs/2406.07514} {arXiv:2406.07514 [physics.ins-det]} \BibitemShut {NoStop}%
\bibitem [{\citenamefont {Cavanna}\ \emph {et~al.}(2014)\citenamefont {Cavanna}, \citenamefont {Kordosky}, \citenamefont {Raaf},\ and\ \citenamefont {Rebel}}]{Cavanna:2014iqa}%
  \BibitemOpen
  \bibfield  {author} {\bibinfo {author} {\bibfnamefont {F.}~\bibnamefont {Cavanna}}, \bibinfo {author} {\bibfnamefont {M.}~\bibnamefont {Kordosky}}, \bibinfo {author} {\bibfnamefont {J.}~\bibnamefont {Raaf}},\ and\ \bibinfo {author} {\bibfnamefont {B.}~\bibnamefont {Rebel}} (\bibinfo {collaboration} {LArIAT}),\ }\href@noop {} {\bibinfo {title} {{LArIAT: Liquid Argon In A Testbeam}}} (\bibinfo {year} {2014}),\ \Eprint {https://arxiv.org/abs/1406.5560} {arXiv:1406.5560 [physics.ins-det]} \BibitemShut {NoStop}%
\bibitem [{\citenamefont {Antonello}\ \emph {et~al.}(2013)\citenamefont {Antonello} \emph {et~al.}}]{Antonello:2013ypa}%
  \BibitemOpen
  \bibfield  {author} {\bibinfo {author} {\bibfnamefont {M.}~\bibnamefont {Antonello}} \emph {et~al.},\ }\href@noop {} {\bibinfo {title} {{ICARUS at FNAL}}} (\bibinfo {year} {2013}),\ \Eprint {https://arxiv.org/abs/1312.7252} {arXiv:1312.7252 [physics.ins-det]} \BibitemShut {NoStop}%
\bibitem [{\citenamefont {Acciarri}\ \emph {et~al.}(2017)\citenamefont {Acciarri} \emph {et~al.}}]{MicroBooNE:2016pwy}%
  \BibitemOpen
  \bibfield  {author} {\bibinfo {author} {\bibfnamefont {R.}~\bibnamefont {Acciarri}} \emph {et~al.} (\bibinfo {collaboration} {MicroBooNE}),\ }\href {https://doi.org/10.1088/1748-0221/12/02/P02017} {\bibfield  {journal} {\bibinfo  {journal} {JINST}\ }\textbf {\bibinfo {volume} {12}}\bibfield  {number} {\bibinfo  {number} { (02)},\ \bibinfo {pages} {P02017}},\ }\Eprint {https://arxiv.org/abs/1612.05824} {arXiv:1612.05824 [physics.ins-det]} \BibitemShut {NoStop}%
\bibitem [{\citenamefont {Acciarri}\ \emph {et~al.}(2020)\citenamefont {Acciarri} \emph {et~al.}}]{SBND:2020scp}%
  \BibitemOpen
  \bibfield  {author} {\bibinfo {author} {\bibfnamefont {R.}~\bibnamefont {Acciarri}} \emph {et~al.} (\bibinfo {collaboration} {SBND}),\ }\href {https://doi.org/10.1088/1748-0221/15/06/P06033} {\bibfield  {journal} {\bibinfo  {journal} {JINST}\ }\textbf {\bibinfo {volume} {15}}\bibfield  {number} {\bibinfo  {number} { (06)},\ \bibinfo {pages} {P06033}},\ }\Eprint {https://arxiv.org/abs/2002.08424} {arXiv:2002.08424 [physics.ins-det]} \BibitemShut {NoStop}%
\bibitem [{\citenamefont {Adhikari}\ \emph {et~al.}(2025)\citenamefont {Adhikari} \emph {et~al.}}]{DEAP:2025shk}%
  \BibitemOpen
  \bibfield  {author} {\bibinfo {author} {\bibfnamefont {P.}~\bibnamefont {Adhikari}} \emph {et~al.} (\bibinfo {collaboration} {DEAP}),\ }\href@noop {} {\bibinfo {title} {{Direct Measurement of the $^{39}$Ar Half-life from 3.4 Years of Data with the DEAP-3600 Detector}}} (\bibinfo {year} {2025}),\ \Eprint {https://arxiv.org/abs/2501.13196} {arXiv:2501.13196 [nucl-ex]} \BibitemShut {NoStop}%
\bibitem [{\citenamefont {Amaudruz}\ \emph {et~al.}(2016)\citenamefont {Amaudruz} \emph {et~al.}}]{DEAP:2014pug}%
  \BibitemOpen
  \bibfield  {author} {\bibinfo {author} {\bibfnamefont {P.~A.}\ \bibnamefont {Amaudruz}} \emph {et~al.} (\bibinfo {collaboration} {DEAP}),\ }\href {https://doi.org/10.1016/j.nuclphysbps.2015.09.048} {\bibfield  {journal} {\bibinfo  {journal} {Nucl. Part. Phys. Proc.}\ }\textbf {\bibinfo {volume} {273-275}},\ \bibinfo {pages} {340} (\bibinfo {year} {2016})},\ \Eprint {https://arxiv.org/abs/1410.7673} {arXiv:1410.7673 [physics.ins-det]} \BibitemShut {NoStop}%
\bibitem [{\citenamefont {Aguilar-Arevalo}\ \emph {et~al.}(2022{\natexlab{a}})\citenamefont {Aguilar-Arevalo} \emph {et~al.}}]{CCM:2021yzc}%
  \BibitemOpen
  \bibfield  {author} {\bibinfo {author} {\bibfnamefont {A.~A.}\ \bibnamefont {Aguilar-Arevalo}} \emph {et~al.} (\bibinfo {collaboration} {CCM}),\ }\href {https://doi.org/10.1103/PhysRevLett.129.021801} {\bibfield  {journal} {\bibinfo  {journal} {Phys. Rev. Lett.}\ }\textbf {\bibinfo {volume} {129}},\ \bibinfo {pages} {021801} (\bibinfo {year} {2022}{\natexlab{a}})},\ \Eprint {https://arxiv.org/abs/2109.14146} {arXiv:2109.14146 [hep-ex]} \BibitemShut {NoStop}%
\bibitem [{\citenamefont {Aguilar-Arevalo}\ \emph {et~al.}(2023)\citenamefont {Aguilar-Arevalo} \emph {et~al.}}]{CCM:2021jmk}%
  \BibitemOpen
  \bibfield  {author} {\bibinfo {author} {\bibfnamefont {A.~A.}\ \bibnamefont {Aguilar-Arevalo}} \emph {et~al.} (\bibinfo {collaboration} {CCM}),\ }\href {https://doi.org/10.1103/PhysRevD.107.095036} {\bibfield  {journal} {\bibinfo  {journal} {Phys. Rev. D}\ }\textbf {\bibinfo {volume} {107}},\ \bibinfo {pages} {095036} (\bibinfo {year} {2023})},\ \Eprint {https://arxiv.org/abs/2112.09979} {arXiv:2112.09979 [hep-ph]} \BibitemShut {NoStop}%
\bibitem [{\citenamefont {Aguilar-Arevalo}\ \emph {et~al.}(2022{\natexlab{b}})\citenamefont {Aguilar-Arevalo} \emph {et~al.}}]{CCM:2021leg}%
  \BibitemOpen
  \bibfield  {author} {\bibinfo {author} {\bibfnamefont {A.~A.}\ \bibnamefont {Aguilar-Arevalo}} \emph {et~al.} (\bibinfo {collaboration} {CCM}),\ }\href {https://doi.org/10.1103/PhysRevD.106.012001} {\bibfield  {journal} {\bibinfo  {journal} {Phys. Rev. D}\ }\textbf {\bibinfo {volume} {106}},\ \bibinfo {pages} {012001} (\bibinfo {year} {2022}{\natexlab{b}})},\ \Eprint {https://arxiv.org/abs/2105.14020} {arXiv:2105.14020 [hep-ex]} \BibitemShut {NoStop}%
\bibitem [{\citenamefont {Aguilar-Arevalo}\ \emph {et~al.}(2024)\citenamefont {Aguilar-Arevalo} \emph {et~al.}}]{CCM:2023itc}%
  \BibitemOpen
  \bibfield  {author} {\bibinfo {author} {\bibfnamefont {A.~A.}\ \bibnamefont {Aguilar-Arevalo}} \emph {et~al.} (\bibinfo {collaboration} {CCM}),\ }\href {https://doi.org/10.1103/PhysRevD.109.095017} {\bibfield  {journal} {\bibinfo  {journal} {Phys. Rev. D}\ }\textbf {\bibinfo {volume} {109}},\ \bibinfo {pages} {095017} (\bibinfo {year} {2024})},\ \bibinfo {note} {[Addendum: Phys.Rev.D 111, 035030 (2025)]},\ \Eprint {https://arxiv.org/abs/2309.02599} {arXiv:2309.02599 [hep-ph]} \BibitemShut {NoStop}%
\bibitem [{\citenamefont {Tripathi}(2024)}]{Tripathi:2024jnq}%
  \BibitemOpen
  \bibfield  {author} {\bibinfo {author} {\bibfnamefont {M.}~\bibnamefont {Tripathi}},\ }\emph {\bibinfo {title} {{Dark Matter Searches at Coherent Captain-Mills}}},\ \href@noop {} {Ph.D. thesis},\ \bibinfo  {school} {Florida U.} (\bibinfo {year} {2024})\BibitemShut {NoStop}%
\bibitem [{\citenamefont {Dunton}(2022)}]{Dunton:2022dez}%
  \BibitemOpen
  \bibfield  {author} {\bibinfo {author} {\bibfnamefont {E.~C.}\ \bibnamefont {Dunton}},\ }\emph {\bibinfo {title} {{A Search for Axion-like Particles at the Coherent CAPTAIN Mills Experiment}}},\ \href {https://doi.org/10.7916/x9x1-ka48} {Ph.D. thesis},\ \bibinfo  {school} {Columbia U.} (\bibinfo {year} {2022})\BibitemShut {NoStop}%
\bibitem [{\citenamefont {Alimonti}\ \emph {et~al.}(2002)\citenamefont {Alimonti} \emph {et~al.}}]{Borexino:2000uvj}%
  \BibitemOpen
  \bibfield  {author} {\bibinfo {author} {\bibfnamefont {G.}~\bibnamefont {Alimonti}} \emph {et~al.} (\bibinfo {collaboration} {Borexino}),\ }\href {https://doi.org/10.1016/S0927-6505(01)00110-4} {\bibfield  {journal} {\bibinfo  {journal} {Astropart. Phys.}\ }\textbf {\bibinfo {volume} {16}},\ \bibinfo {pages} {205} (\bibinfo {year} {2002})},\ \Eprint {https://arxiv.org/abs/hep-ex/0012030} {arXiv:hep-ex/0012030} \BibitemShut {NoStop}%
\bibitem [{\citenamefont {Andringa}\ \emph {et~al.}(2016)\citenamefont {Andringa} \emph {et~al.}}]{SNO:2015wyx}%
  \BibitemOpen
  \bibfield  {author} {\bibinfo {author} {\bibfnamefont {S.}~\bibnamefont {Andringa}} \emph {et~al.} (\bibinfo {collaboration} {SNO+}),\ }\href {https://doi.org/10.1155/2016/6194250} {\bibfield  {journal} {\bibinfo  {journal} {Adv. High Energy Phys.}\ }\textbf {\bibinfo {volume} {2016}},\ \bibinfo {pages} {6194250} (\bibinfo {year} {2016})},\ \Eprint {https://arxiv.org/abs/1508.05759} {arXiv:1508.05759 [physics.ins-det]} \BibitemShut {NoStop}%
\bibitem [{\citenamefont {Aguilar-Arevalo}\ \emph {et~al.}(2025)\citenamefont {Aguilar-Arevalo} \emph {et~al.}}]{companion_paper}%
  \BibitemOpen
  \bibfield  {author} {\bibinfo {author} {\bibfnamefont {A.~A.}\ \bibnamefont {Aguilar-Arevalo}} \emph {et~al.},\ }\href@noop {} {\bibinfo {title} {{First Event-by-Event Identification of Cherenkov Radiation from Sub-MeV Particles in Liquid Argon}}} (\bibinfo {year} {2025}),\ \Eprint {https://arxiv.org/abs/2507.08886} {arXiv:2507.08886 [physics.ins-det]} \BibitemShut {NoStop}%
\bibitem [{\citenamefont {Rahman}(2024)}]{Rahman:2024zhp}%
  \BibitemOpen
  \bibfield  {author} {\bibinfo {author} {\bibfnamefont {H.~R.}\ \bibnamefont {Rahman}},\ }\emph {\bibinfo {title} {{Calculating the Total Cherenkov Radiation Emitted by Low Energy Protons in Liquid Argon and Comparing with Argon Scintillation Light at 128 nm}}},\ \href@noop {} {Master's thesis},\ \bibinfo  {school} {New Mexico State U.} (\bibinfo {year} {2024}),\ \Eprint {https://arxiv.org/abs/2408.00817} {arXiv:2408.00817 [hep-ph]} \BibitemShut {NoStop}%
\bibitem [{\citenamefont {Born}\ and\ \citenamefont {Wolf}(1999)}]{born1999principles}%
  \BibitemOpen
  \bibfield  {author} {\bibinfo {author} {\bibfnamefont {M.}~\bibnamefont {Born}}\ and\ \bibinfo {author} {\bibfnamefont {E.}~\bibnamefont {Wolf}},\ }\href@noop {} {\emph {\bibinfo {title} {Principles of Optics}}},\ \bibinfo {edition} {7th}\ ed.\ (\bibinfo  {publisher} {Cambridge University Press},\ \bibinfo {year} {1999})\BibitemShut {NoStop}%
\bibitem [{\citenamefont {Babicz}\ \emph {et~al.}(2020)\citenamefont {Babicz} \emph {et~al.}}]{Babicz:2020den}%
  \BibitemOpen
  \bibfield  {author} {\bibinfo {author} {\bibfnamefont {M.}~\bibnamefont {Babicz}} \emph {et~al.},\ }\href {https://doi.org/10.1088/1748-0221/15/09/P09009} {\bibfield  {journal} {\bibinfo  {journal} {JINST}\ }\textbf {\bibinfo {volume} {15}}\bibfield  {number} {\bibinfo  {number} { (09)},\ \bibinfo {pages} {P09009}},\ }\Eprint {https://arxiv.org/abs/2002.09346} {arXiv:2002.09346 [physics.ins-det]} \BibitemShut {NoStop}%
\bibitem [{\citenamefont {Grace}\ and\ \citenamefont {Nikkel}(2017)}]{Grace:2015yta}%
  \BibitemOpen
  \bibfield  {author} {\bibinfo {author} {\bibfnamefont {E.}~\bibnamefont {Grace}}\ and\ \bibinfo {author} {\bibfnamefont {J.~A.}\ \bibnamefont {Nikkel}},\ }\href {https://doi.org/10.1016/j.nima.2017.06.031} {\bibfield  {journal} {\bibinfo  {journal} {Nucl. Instrum. Meth. A}\ }\textbf {\bibinfo {volume} {867}},\ \bibinfo {pages} {204} (\bibinfo {year} {2017})},\ \Eprint {https://arxiv.org/abs/1502.04213} {arXiv:1502.04213 [physics.ins-det]} \BibitemShut {NoStop}%
\bibitem [{\citenamefont {Degrave}\ \emph {et~al.}(2016)\citenamefont {Degrave}, \citenamefont {Hermans}, \citenamefont {Dambre},\ and\ \citenamefont {wyffels}}]{degrave2016differentiable}%
  \BibitemOpen
  \bibfield  {author} {\bibinfo {author} {\bibfnamefont {J.}~\bibnamefont {Degrave}}, \bibinfo {author} {\bibfnamefont {M.}~\bibnamefont {Hermans}}, \bibinfo {author} {\bibfnamefont {J.}~\bibnamefont {Dambre}},\ and\ \bibinfo {author} {\bibfnamefont {F.}~\bibnamefont {wyffels}},\ }\href {https://arxiv.org/abs/1611.01652} {\bibinfo {title} {A differentiable physics engine for deep learning in robotics}} (\bibinfo {year} {2016}),\ \Eprint {https://arxiv.org/abs/1611.01652} {arXiv:1611.01652 [cs.LG]} \BibitemShut {NoStop}%
\bibitem [{\citenamefont {Todorov}\ \emph {et~al.}(2012)\citenamefont {Todorov}, \citenamefont {Erez},\ and\ \citenamefont {Tassa}}]{6386109}%
  \BibitemOpen
  \bibfield  {author} {\bibinfo {author} {\bibfnamefont {E.}~\bibnamefont {Todorov}}, \bibinfo {author} {\bibfnamefont {T.}~\bibnamefont {Erez}},\ and\ \bibinfo {author} {\bibfnamefont {Y.}~\bibnamefont {Tassa}},\ }in\ \href {https://doi.org/10.1109/IROS.2012.6386109} {\emph {\bibinfo {booktitle} {2012 IEEE/RSJ International Conference on Intelligent Robots and Systems}}}\ (\bibinfo {year} {2012})\ pp.\ \bibinfo {pages} {5026--5033}\BibitemShut {NoStop}%
\bibitem [{\citenamefont {Newbury}\ \emph {et~al.}(2024)\citenamefont {Newbury}, \citenamefont {Collins}, \citenamefont {He}, \citenamefont {Pan}, \citenamefont {Posner}, \citenamefont {Howard},\ and\ \citenamefont {Cosgun}}]{newbury2024reviewdifferentiablesimulators}%
  \BibitemOpen
  \bibfield  {author} {\bibinfo {author} {\bibfnamefont {R.}~\bibnamefont {Newbury}}, \bibinfo {author} {\bibfnamefont {J.}~\bibnamefont {Collins}}, \bibinfo {author} {\bibfnamefont {K.}~\bibnamefont {He}}, \bibinfo {author} {\bibfnamefont {J.}~\bibnamefont {Pan}}, \bibinfo {author} {\bibfnamefont {I.}~\bibnamefont {Posner}}, \bibinfo {author} {\bibfnamefont {D.}~\bibnamefont {Howard}},\ and\ \bibinfo {author} {\bibfnamefont {A.}~\bibnamefont {Cosgun}},\ }\href {https://arxiv.org/abs/2407.05560} {\bibinfo {title} {A review of differentiable simulators}} (\bibinfo {year} {2024}),\ \Eprint {https://arxiv.org/abs/2407.05560} {arXiv:2407.05560 [cs.RO]} \BibitemShut {NoStop}%
\bibitem [{\citenamefont {Byrd}\ \emph {et~al.}(1995)\citenamefont {Byrd}, \citenamefont {Lu}, \citenamefont {Nocedal},\ and\ \citenamefont {Zhu}}]{minimizer1}%
  \BibitemOpen
  \bibfield  {author} {\bibinfo {author} {\bibfnamefont {R.~H.}\ \bibnamefont {Byrd}}, \bibinfo {author} {\bibfnamefont {P.}~\bibnamefont {Lu}}, \bibinfo {author} {\bibfnamefont {J.}~\bibnamefont {Nocedal}},\ and\ \bibinfo {author} {\bibfnamefont {C.}~\bibnamefont {Zhu}},\ }\href {https://doi.org/10.1137/0916069} {\bibfield  {journal} {\bibinfo  {journal} {SIAM Journal on Scientific Computing}\ }\textbf {\bibinfo {volume} {16}},\ \bibinfo {pages} {1190} (\bibinfo {year} {1995})},\ \Eprint {https://arxiv.org/abs/https://doi.org/10.1137/0916069} {https://doi.org/10.1137/0916069} \BibitemShut {NoStop}%
\bibitem [{\citenamefont {Zhu}\ \emph {et~al.}(1997)\citenamefont {Zhu}, \citenamefont {Byrd}, \citenamefont {Lu},\ and\ \citenamefont {Nocedal}}]{minimizer2}%
  \BibitemOpen
  \bibfield  {author} {\bibinfo {author} {\bibfnamefont {C.}~\bibnamefont {Zhu}}, \bibinfo {author} {\bibfnamefont {R.~H.}\ \bibnamefont {Byrd}}, \bibinfo {author} {\bibfnamefont {P.}~\bibnamefont {Lu}},\ and\ \bibinfo {author} {\bibfnamefont {J.}~\bibnamefont {Nocedal}},\ }\href {https://doi.org/10.1145/279232.279236} {\bibfield  {journal} {\bibinfo  {journal} {ACM Trans. Math. Softw.}\ }\textbf {\bibinfo {volume} {23}},\ \bibinfo {pages} {550–560} (\bibinfo {year} {1997})}\BibitemShut {NoStop}%
\bibitem [{\citenamefont {Doke}\ \emph {et~al.}(1990)\citenamefont {Doke}, \citenamefont {Masuda},\ and\ \citenamefont {Shibamura}}]{Doke:1990rza}%
  \BibitemOpen
  \bibfield  {author} {\bibinfo {author} {\bibfnamefont {T.}~\bibnamefont {Doke}}, \bibinfo {author} {\bibfnamefont {K.}~\bibnamefont {Masuda}},\ and\ \bibinfo {author} {\bibfnamefont {E.}~\bibnamefont {Shibamura}},\ }\href {https://doi.org/10.1016/0168-9002(90)90011-T} {\bibfield  {journal} {\bibinfo  {journal} {Nucl. Instrum. Meth. A}\ }\textbf {\bibinfo {volume} {291}},\ \bibinfo {pages} {617} (\bibinfo {year} {1990})}\BibitemShut {NoStop}%
\bibitem [{\citenamefont {Whittington}\ \emph {et~al.}(2016)\citenamefont {Whittington}, \citenamefont {Mufson},\ and\ \citenamefont {Howard}}]{Whittington:2014aha}%
  \BibitemOpen
  \bibfield  {author} {\bibinfo {author} {\bibfnamefont {D.}~\bibnamefont {Whittington}}, \bibinfo {author} {\bibfnamefont {S.}~\bibnamefont {Mufson}},\ and\ \bibinfo {author} {\bibfnamefont {B.}~\bibnamefont {Howard}},\ }\href {https://doi.org/10.1088/1748-0221/11/05/P05016} {\bibfield  {journal} {\bibinfo  {journal} {JINST}\ }\textbf {\bibinfo {volume} {11}}\bibfield  {number} {\bibinfo  {number} { (05)},\ \bibinfo {pages} {P05016}},\ }\Eprint {https://arxiv.org/abs/1408.1763} {arXiv:1408.1763 [physics.ins-det]} \BibitemShut {NoStop}%
\bibitem [{\citenamefont {Segreto}(2021)}]{Segreto:2020qks}%
  \BibitemOpen
  \bibfield  {author} {\bibinfo {author} {\bibfnamefont {E.}~\bibnamefont {Segreto}},\ }\href {https://doi.org/10.1103/PhysRevD.103.043001} {\bibfield  {journal} {\bibinfo  {journal} {Phys. Rev. D}\ }\textbf {\bibinfo {volume} {103}},\ \bibinfo {pages} {043001} (\bibinfo {year} {2021})},\ \Eprint {https://arxiv.org/abs/2012.06527} {arXiv:2012.06527 [physics.ins-det]} \BibitemShut {NoStop}%
\bibitem [{\citenamefont {Adhikari}\ \emph {et~al.}(2020)\citenamefont {Adhikari} \emph {et~al.}}]{DEAP:2020hms}%
  \BibitemOpen
  \bibfield  {author} {\bibinfo {author} {\bibfnamefont {P.}~\bibnamefont {Adhikari}} \emph {et~al.} (\bibinfo {collaboration} {DEAP}),\ }\href {https://doi.org/10.1140/epjc/s10052-020-7789-x} {\bibfield  {journal} {\bibinfo  {journal} {Eur. Phys. J. C}\ }\textbf {\bibinfo {volume} {80}},\ \bibinfo {pages} {303} (\bibinfo {year} {2020})},\ \Eprint {https://arxiv.org/abs/2001.09855} {arXiv:2001.09855 [physics.ins-det]} \BibitemShut {NoStop}%
\bibitem [{\citenamefont {Hofmann}\ \emph {et~al.}(2013)\citenamefont {Hofmann}, \citenamefont {Dandl}, \citenamefont {Heindl}, \citenamefont {Neumeier}, \citenamefont {Oberauer}, \citenamefont {Potzel}, \citenamefont {Roth}, \citenamefont {Sch\"oNert}, \citenamefont {Wieser},\ and\ \citenamefont {Ulrich}}]{Hofmann:2013vva}%
  \BibitemOpen
  \bibfield  {author} {\bibinfo {author} {\bibfnamefont {M.}~\bibnamefont {Hofmann}}, \bibinfo {author} {\bibfnamefont {T.}~\bibnamefont {Dandl}}, \bibinfo {author} {\bibfnamefont {T.}~\bibnamefont {Heindl}}, \bibinfo {author} {\bibfnamefont {A.}~\bibnamefont {Neumeier}}, \bibinfo {author} {\bibfnamefont {L.}~\bibnamefont {Oberauer}}, \bibinfo {author} {\bibfnamefont {W.}~\bibnamefont {Potzel}}, \bibinfo {author} {\bibfnamefont {S.}~\bibnamefont {Roth}}, \bibinfo {author} {\bibfnamefont {S.}~\bibnamefont {Sch\"oNert}}, \bibinfo {author} {\bibfnamefont {J.}~\bibnamefont {Wieser}},\ and\ \bibinfo {author} {\bibfnamefont {A.}~\bibnamefont {Ulrich}},\ }\href {https://doi.org/10.1140/epjc/s10052-013-2618-0} {\bibfield  {journal} {\bibinfo  {journal} {Eur. Phys. J. C}\ }\textbf {\bibinfo {volume} {73}},\ \bibinfo {pages} {2618} (\bibinfo {year} {2013})},\ \Eprint {https://arxiv.org/abs/1511.07721} {arXiv:1511.07721 [physics.ins-det]} \BibitemShut {NoStop}%
\bibitem [{\citenamefont {Heindl}\ \emph {et~al.}(2010)\citenamefont {Heindl}, \citenamefont {Dandl}, \citenamefont {Hofmann}, \citenamefont {Krucken}, \citenamefont {Oberauer}, \citenamefont {Potzel}, \citenamefont {Wieser},\ and\ \citenamefont {Ulrich}}]{Heindl:2010zz}%
  \BibitemOpen
  \bibfield  {author} {\bibinfo {author} {\bibfnamefont {T.}~\bibnamefont {Heindl}}, \bibinfo {author} {\bibfnamefont {T.}~\bibnamefont {Dandl}}, \bibinfo {author} {\bibfnamefont {M.}~\bibnamefont {Hofmann}}, \bibinfo {author} {\bibfnamefont {R.}~\bibnamefont {Krucken}}, \bibinfo {author} {\bibfnamefont {L.}~\bibnamefont {Oberauer}}, \bibinfo {author} {\bibfnamefont {W.}~\bibnamefont {Potzel}}, \bibinfo {author} {\bibfnamefont {J.}~\bibnamefont {Wieser}},\ and\ \bibinfo {author} {\bibfnamefont {A.}~\bibnamefont {Ulrich}},\ }\href {https://doi.org/10.1209/0295-5075/91/62002} {\bibfield  {journal} {\bibinfo  {journal} {EPL}\ }\textbf {\bibinfo {volume} {91}},\ \bibinfo {pages} {62002} (\bibinfo {year} {2010})},\ \Eprint {https://arxiv.org/abs/1511.07718} {arXiv:1511.07718 [physics.ins-det]} \BibitemShut {NoStop}%
\bibitem [{\citenamefont {Kr\"otz}\ \emph {et~al.}(1991)\citenamefont {Kr\"otz}, \citenamefont {Ulrich}, \citenamefont {Busch}, \citenamefont {Ribitzki},\ and\ \citenamefont {Wieser}}]{PhysRevA.43.6089}%
  \BibitemOpen
  \bibfield  {author} {\bibinfo {author} {\bibfnamefont {W.}~\bibnamefont {Kr\"otz}}, \bibinfo {author} {\bibfnamefont {A.}~\bibnamefont {Ulrich}}, \bibinfo {author} {\bibfnamefont {B.}~\bibnamefont {Busch}}, \bibinfo {author} {\bibfnamefont {G.}~\bibnamefont {Ribitzki}},\ and\ \bibinfo {author} {\bibfnamefont {J.}~\bibnamefont {Wieser}},\ }\href {https://doi.org/10.1103/PhysRevA.43.6089} {\bibfield  {journal} {\bibinfo  {journal} {Phys. Rev. A}\ }\textbf {\bibinfo {volume} {43}},\ \bibinfo {pages} {6089} (\bibinfo {year} {1991})}\BibitemShut {NoStop}%
\bibitem [{\citenamefont {Benson}\ \emph {et~al.}(2018)\citenamefont {Benson}, \citenamefont {Orebi~Gann},\ and\ \citenamefont {Gehman}}]{Benson:2017vbw}%
  \BibitemOpen
  \bibfield  {author} {\bibinfo {author} {\bibfnamefont {C.}~\bibnamefont {Benson}}, \bibinfo {author} {\bibfnamefont {G.}~\bibnamefont {Orebi~Gann}},\ and\ \bibinfo {author} {\bibfnamefont {V.}~\bibnamefont {Gehman}},\ }\href {https://doi.org/10.1140/s10052-018-5807-z} {\bibfield  {journal} {\bibinfo  {journal} {Eur. Phys. J. C}\ }\textbf {\bibinfo {volume} {78}},\ \bibinfo {pages} {329} (\bibinfo {year} {2018})},\ \Eprint {https://arxiv.org/abs/1709.05002} {arXiv:1709.05002 [physics.ins-det]} \BibitemShut {NoStop}%
\bibitem [{\citenamefont {Wallace-Williams}\ \emph {et~al.}(1994)\citenamefont {Wallace-Williams}, \citenamefont {Schwartz}, \citenamefont {Moeller}, \citenamefont {Goldbeck}, \citenamefont {Yee}, \citenamefont {El-Bayoumi},\ and\ \citenamefont {Kliger}}]{doi:10.1021/j100052a011}%
  \BibitemOpen
  \bibfield  {author} {\bibinfo {author} {\bibfnamefont {S.~E.}\ \bibnamefont {Wallace-Williams}}, \bibinfo {author} {\bibfnamefont {B.~J.}\ \bibnamefont {Schwartz}}, \bibinfo {author} {\bibfnamefont {S.}~\bibnamefont {Moeller}}, \bibinfo {author} {\bibfnamefont {R.~A.}\ \bibnamefont {Goldbeck}}, \bibinfo {author} {\bibfnamefont {W.~A.}\ \bibnamefont {Yee}}, \bibinfo {author} {\bibfnamefont {M.~A.}\ \bibnamefont {El-Bayoumi}},\ and\ \bibinfo {author} {\bibfnamefont {D.~S.}\ \bibnamefont {Kliger}},\ }\href {https://doi.org/10.1021/j100052a011} {\bibfield  {journal} {\bibinfo  {journal} {The Journal of Physical Chemistry}\ }\textbf {\bibinfo {volume} {98}},\ \bibinfo {pages} {60} (\bibinfo {year} {1994})},\ \Eprint {https://arxiv.org/abs/https://doi.org/10.1021/j100052a011} {https://doi.org/10.1021/j100052a011} \BibitemShut {NoStop}%
\bibitem [{\citenamefont {Acciarri}\ \emph {et~al.}(2010{\natexlab{a}})\citenamefont {Acciarri} \emph {et~al.}}]{WArP:2008rgv}%
  \BibitemOpen
  \bibfield  {author} {\bibinfo {author} {\bibfnamefont {R.}~\bibnamefont {Acciarri}} \emph {et~al.} (\bibinfo {collaboration} {WArP}),\ }\href {https://doi.org/10.1088/1748-0221/5/06/P06003} {\bibfield  {journal} {\bibinfo  {journal} {JINST}\ }\textbf {\bibinfo {volume} {5}},\ \bibinfo {pages} {P06003}},\ \Eprint {https://arxiv.org/abs/0804.1217} {arXiv:0804.1217 [nucl-ex]} \BibitemShut {NoStop}%
\bibitem [{Mic(2022)}]{MicroBooNE:2022pcx}%
  \BibitemOpen
  \href {https://doi.org/10.2172/2397218} {\bibinfo {title} {{Light Yield Calibration in MicroBooNE}}} (\bibinfo {year} {2022})\BibitemShut {NoStop}%
\bibitem [{\citenamefont {Mavrokoridis}\ \emph {et~al.}(2011)\citenamefont {Mavrokoridis}, \citenamefont {Calland}, \citenamefont {Coleman}, \citenamefont {Lightfoot}, \citenamefont {McCauley}, \citenamefont {McCormick},\ and\ \citenamefont {Touramanis}}]{Mavrokoridis:2011wv}%
  \BibitemOpen
  \bibfield  {author} {\bibinfo {author} {\bibfnamefont {K.}~\bibnamefont {Mavrokoridis}}, \bibinfo {author} {\bibfnamefont {R.~G.}\ \bibnamefont {Calland}}, \bibinfo {author} {\bibfnamefont {J.}~\bibnamefont {Coleman}}, \bibinfo {author} {\bibfnamefont {P.~K.}\ \bibnamefont {Lightfoot}}, \bibinfo {author} {\bibfnamefont {N.}~\bibnamefont {McCauley}}, \bibinfo {author} {\bibfnamefont {K.~J.}\ \bibnamefont {McCormick}},\ and\ \bibinfo {author} {\bibfnamefont {C.}~\bibnamefont {Touramanis}},\ }\href {https://doi.org/10.1088/1748-0221/6/08/P08003} {\bibfield  {journal} {\bibinfo  {journal} {JINST}\ }\textbf {\bibinfo {volume} {6}},\ \bibinfo {pages} {P08003}},\ \Eprint {https://arxiv.org/abs/1106.5226} {arXiv:1106.5226 [physics.ins-det]} \BibitemShut {NoStop}%
\bibitem [{\citenamefont {Cherenkov}(1937)}]{Cherenkov:1937mnd}%
  \BibitemOpen
  \bibfield  {author} {\bibinfo {author} {\bibfnamefont {P.~A.}\ \bibnamefont {Cherenkov}},\ }\href {https://doi.org/10.1103/PhysRev.52.378} {\bibfield  {journal} {\bibinfo  {journal} {Phys. Rev.}\ }\textbf {\bibinfo {volume} {52}},\ \bibinfo {pages} {378} (\bibinfo {year} {1937})}\BibitemShut {NoStop}%
\bibitem [{\citenamefont {Patterson}\ \emph {et~al.}(2009)\citenamefont {Patterson}, \citenamefont {Laird}, \citenamefont {Liu}, \citenamefont {Meyers}, \citenamefont {Stancu},\ and\ \citenamefont {Tanaka}}]{Patterson:2009ki}%
  \BibitemOpen
  \bibfield  {author} {\bibinfo {author} {\bibfnamefont {R.~B.}\ \bibnamefont {Patterson}}, \bibinfo {author} {\bibfnamefont {E.~M.}\ \bibnamefont {Laird}}, \bibinfo {author} {\bibfnamefont {Y.}~\bibnamefont {Liu}}, \bibinfo {author} {\bibfnamefont {P.~D.}\ \bibnamefont {Meyers}}, \bibinfo {author} {\bibfnamefont {I.}~\bibnamefont {Stancu}},\ and\ \bibinfo {author} {\bibfnamefont {H.~A.}\ \bibnamefont {Tanaka}},\ }\href {https://doi.org/10.1016/j.nima.2009.06.064} {\bibfield  {journal} {\bibinfo  {journal} {Nucl. Instrum. Meth. A}\ }\textbf {\bibinfo {volume} {608}},\ \bibinfo {pages} {206} (\bibinfo {year} {2009})},\ \Eprint {https://arxiv.org/abs/0902.2222} {arXiv:0902.2222 [hep-ex]} \BibitemShut {NoStop}%
\bibitem [{\citenamefont {Jiang}\ \emph {et~al.}(2019)\citenamefont {Jiang} \emph {et~al.}}]{Super-Kamiokande:2019gzr}%
  \BibitemOpen
  \bibfield  {author} {\bibinfo {author} {\bibfnamefont {M.}~\bibnamefont {Jiang}} \emph {et~al.} (\bibinfo {collaboration} {Super-Kamiokande}),\ }\href {https://doi.org/10.1093/ptep/ptz015} {\bibfield  {journal} {\bibinfo  {journal} {PTEP}\ }\textbf {\bibinfo {volume} {2019}},\ \bibinfo {pages} {053F01} (\bibinfo {year} {2019})},\ \Eprint {https://arxiv.org/abs/1901.03230} {arXiv:1901.03230 [hep-ex]} \BibitemShut {NoStop}%
\bibitem [{\citenamefont {Group}\ and\ \citenamefont {Workman}(2022)}]{10.1093/ptep/ptac097}%
  \BibitemOpen
  \bibfield  {author} {\bibinfo {author} {\bibfnamefont {P.~D.}\ \bibnamefont {Group}}\ and\ \bibinfo {author} {\bibnamefont {Workman}},\ }\href {https://doi.org/10.1093/ptep/ptac097} {\bibfield  {journal} {\bibinfo  {journal} {Progress of Theoretical and Experimental Physics}\ }\textbf {\bibinfo {volume} {2022}},\ \bibinfo {pages} {083C01} (\bibinfo {year} {2022})},\ \Eprint {https://arxiv.org/abs/https://academic.oup.com/ptep/article-pdf/2022/8/083C01/49175539/ptac097.pdf} {https://academic.oup.com/ptep/article-pdf/2022/8/083C01/49175539/ptac097.pdf} \BibitemShut {NoStop}%
\bibitem [{\citenamefont {Sinnock}\ and\ \citenamefont {Smith}(1969)}]{Sinnock:1969zz}%
  \BibitemOpen
  \bibfield  {author} {\bibinfo {author} {\bibfnamefont {A.~C.}\ \bibnamefont {Sinnock}}\ and\ \bibinfo {author} {\bibfnamefont {B.~L.}\ \bibnamefont {Smith}},\ }\href {https://doi.org/10.1103/PhysRev.181.1297} {\bibfield  {journal} {\bibinfo  {journal} {Phys. Rev.}\ }\textbf {\bibinfo {volume} {181}},\ \bibinfo {pages} {1297} (\bibinfo {year} {1969})}\BibitemShut {NoStop}%
\bibitem [{\citenamefont {Lane}\ and\ \citenamefont {Kuppermann}(1968)}]{Lane1968}%
  \BibitemOpen
  \bibfield  {author} {\bibinfo {author} {\bibfnamefont {A.~L.}\ \bibnamefont {Lane}}\ and\ \bibinfo {author} {\bibfnamefont {A.}~\bibnamefont {Kuppermann}},\ }\href {https://doi.org/10.1063/1.1683289} {\bibfield  {journal} {\bibinfo  {journal} {Review of Scientific Instruments}\ }\textbf {\bibinfo {volume} {39}},\ \bibinfo {pages} {126} (\bibinfo {year} {1968})}\BibitemShut {NoStop}%
\bibitem [{\citenamefont {Arai}\ \emph {et~al.}(1978)\citenamefont {Arai}, \citenamefont {Oka}, \citenamefont {Kogoma},\ and\ \citenamefont {Imamura}}]{Arai1978}%
  \BibitemOpen
  \bibfield  {author} {\bibinfo {author} {\bibfnamefont {S.}~\bibnamefont {Arai}}, \bibinfo {author} {\bibfnamefont {T.}~\bibnamefont {Oka}}, \bibinfo {author} {\bibfnamefont {M.}~\bibnamefont {Kogoma}},\ and\ \bibinfo {author} {\bibfnamefont {M.}~\bibnamefont {Imamura}},\ }\href {https://doi.org/10.1063/1.435489} {\bibfield  {journal} {\bibinfo  {journal} {The Journal of Chemical Physics}\ }\textbf {\bibinfo {volume} {68}},\ \bibinfo {pages} {4595} (\bibinfo {year} {1978})}\BibitemShut {NoStop}%
\bibitem [{\citenamefont {Acciarri}\ \emph {et~al.}(2010{\natexlab{b}})\citenamefont {Acciarri} \emph {et~al.}}]{WArP:2008dyo}%
  \BibitemOpen
  \bibfield  {author} {\bibinfo {author} {\bibfnamefont {R.}~\bibnamefont {Acciarri}} \emph {et~al.} (\bibinfo {collaboration} {WArP}),\ }\href {https://doi.org/10.1088/1748-0221/5/05/P05003} {\bibfield  {journal} {\bibinfo  {journal} {JINST}\ }\textbf {\bibinfo {volume} {5}},\ \bibinfo {pages} {P05003}},\ \Eprint {https://arxiv.org/abs/0804.1222} {arXiv:0804.1222 [nucl-ex]} \BibitemShut {NoStop}%
\bibitem [{\citenamefont {Jones}\ \emph {et~al.}(2013)\citenamefont {Jones}, \citenamefont {Chiu}, \citenamefont {Conrad}, \citenamefont {Ignarra}, \citenamefont {Katori},\ and\ \citenamefont {Toups}}]{Jones:2013bca}%
  \BibitemOpen
  \bibfield  {author} {\bibinfo {author} {\bibfnamefont {B.~J.~P.}\ \bibnamefont {Jones}}, \bibinfo {author} {\bibfnamefont {C.~S.}\ \bibnamefont {Chiu}}, \bibinfo {author} {\bibfnamefont {J.~M.}\ \bibnamefont {Conrad}}, \bibinfo {author} {\bibfnamefont {C.~M.}\ \bibnamefont {Ignarra}}, \bibinfo {author} {\bibfnamefont {T.}~\bibnamefont {Katori}},\ and\ \bibinfo {author} {\bibfnamefont {M.}~\bibnamefont {Toups}},\ }\href {https://doi.org/10.1088/1748-0221/8/07/P07011} {\bibfield  {journal} {\bibinfo  {journal} {JINST}\ }\textbf {\bibinfo {volume} {8}},\ \bibinfo {pages} {P07011}},\ \bibinfo {note} {[Erratum: JINST 8, E09001 (2013)]},\ \Eprint {https://arxiv.org/abs/1306.4605} {arXiv:1306.4605 [physics.ins-det]} \BibitemShut {NoStop}%
\bibitem [{\citenamefont {Lawson}\ and\ \citenamefont {Hanson}(1974)}]{lawson1974least}%
  \BibitemOpen
  \bibfield  {author} {\bibinfo {author} {\bibfnamefont {C.~L.}\ \bibnamefont {Lawson}}\ and\ \bibinfo {author} {\bibfnamefont {R.~J.}\ \bibnamefont {Hanson}},\ }\href@noop {} {\emph {\bibinfo {title} {Solving Least Squares Problems}}}\ (\bibinfo  {publisher} {Prentice-Hall},\ \bibinfo {address} {Englewood Cliffs, NJ},\ \bibinfo {year} {1974})\BibitemShut {NoStop}%
\bibitem [{\citenamefont {Aartsen}\ \emph {et~al.}(2014)\citenamefont {Aartsen} \emph {et~al.}}]{IceCube:2013dkx}%
  \BibitemOpen
  \bibfield  {author} {\bibinfo {author} {\bibfnamefont {M.~G.}\ \bibnamefont {Aartsen}} \emph {et~al.} (\bibinfo {collaboration} {IceCube}),\ }\href {https://doi.org/10.1088/1748-0221/9/03/P03009} {\bibfield  {journal} {\bibinfo  {journal} {JINST}\ }\textbf {\bibinfo {volume} {9}},\ \bibinfo {pages} {P03009}},\ \Eprint {https://arxiv.org/abs/1311.4767} {arXiv:1311.4767 [physics.ins-det]} \BibitemShut {NoStop}%
\bibitem [{\citenamefont {Kaptanoglu}(2018)}]{Kaptanoglu:2017jxo}%
  \BibitemOpen
  \bibfield  {author} {\bibinfo {author} {\bibfnamefont {T.}~\bibnamefont {Kaptanoglu}},\ }\href {https://doi.org/10.1016/j.nima.2018.01.086} {\bibfield  {journal} {\bibinfo  {journal} {Nucl. Instrum. Meth. A}\ }\textbf {\bibinfo {volume} {889}},\ \bibinfo {pages} {69} (\bibinfo {year} {2018})},\ \Eprint {https://arxiv.org/abs/1710.03334} {arXiv:1710.03334 [physics.ins-det]} \BibitemShut {NoStop}%
\bibitem [{\citenamefont {Abbasi}\ \emph {et~al.}(2024)\citenamefont {Abbasi} \emph {et~al.}}]{I3DOMCalibration}%
  \BibitemOpen
  \bibfield  {author} {\bibinfo {author} {\bibfnamefont {R.}~\bibnamefont {Abbasi}} \emph {et~al.} (\bibinfo {collaboration} {IceCube Collaboration}),\ }\href@noop {} {\bibinfo {title} {{\texttt{I3DOMCalibration.cxx}}}},\ \bibinfo {howpublished} {\url{https://github.com/icecube/icetray-public/blob/main/dataclasses/private/dataclasses/calibration/I3DOMCalibration.cxx}} (\bibinfo {year} {2024}),\ \bibinfo {note} {accessed: 2025-07-21}\BibitemShut {NoStop}%
\bibitem [{\citenamefont {Abbasi}\ \emph {et~al.}(2009)\citenamefont {Abbasi} \emph {et~al.}}]{IceCube:2008qbc}%
  \BibitemOpen
  \bibfield  {author} {\bibinfo {author} {\bibfnamefont {R.}~\bibnamefont {Abbasi}} \emph {et~al.} (\bibinfo {collaboration} {IceCube}),\ }\href {https://doi.org/10.1016/j.nima.2009.01.001} {\bibfield  {journal} {\bibinfo  {journal} {Nucl. Instrum. Meth. A}\ }\textbf {\bibinfo {volume} {601}},\ \bibinfo {pages} {294} (\bibinfo {year} {2009})},\ \Eprint {https://arxiv.org/abs/0810.4930} {arXiv:0810.4930 [physics.ins-det]} \BibitemShut {NoStop}%
\bibitem [{\citenamefont {Aartsen}\ \emph {et~al.}(2017)\citenamefont {Aartsen} \emph {et~al.}}]{IceCube:2016zyt}%
  \BibitemOpen
  \bibfield  {author} {\bibinfo {author} {\bibfnamefont {M.~G.}\ \bibnamefont {Aartsen}} \emph {et~al.} (\bibinfo {collaboration} {IceCube}),\ }\href {https://doi.org/10.1088/1748-0221/12/03/P03012} {\bibfield  {journal} {\bibinfo  {journal} {JINST}\ }\textbf {\bibinfo {volume} {12}}\bibfield  {number} {\bibinfo  {number} { (03)},\ \bibinfo {pages} {P03012}},\ }\bibinfo {note} {[Erratum: JINST 19, E05001 (2024)]},\ \Eprint {https://arxiv.org/abs/1612.05093} {arXiv:1612.05093 [astro-ph.IM]} \BibitemShut {NoStop}%
\bibitem [{\citenamefont {Caldwell}\ \emph {et~al.}(2013)\citenamefont {Caldwell}, \citenamefont {Seibert},\ and\ \citenamefont {Jaditz}}]{Caldwell:2013oea}%
  \BibitemOpen
  \bibfield  {author} {\bibinfo {author} {\bibfnamefont {T.}~\bibnamefont {Caldwell}}, \bibinfo {author} {\bibfnamefont {S.}~\bibnamefont {Seibert}},\ and\ \bibinfo {author} {\bibfnamefont {S.}~\bibnamefont {Jaditz}},\ }\href {https://doi.org/10.1088/1748-0221/8/09/C09004} {\bibfield  {journal} {\bibinfo  {journal} {JINST}\ }\textbf {\bibinfo {volume} {8}},\ \bibinfo {pages} {C09004}}\BibitemShut {NoStop}%
\bibitem [{\citenamefont {Basunia}(2015)}]{BASUNIA201569}%
  \BibitemOpen
  \bibfield  {author} {\bibinfo {author} {\bibfnamefont {M.~S.}\ \bibnamefont {Basunia}},\ }\href {https://doi.org/https://doi.org/10.1016/j.nds.2015.07.002} {\bibfield  {journal} {\bibinfo  {journal} {Nuclear Data Sheets}\ }\textbf {\bibinfo {volume} {127}},\ \bibinfo {pages} {69} (\bibinfo {year} {2015})}\BibitemShut {NoStop}%
\bibitem [{\citenamefont {Agostinelli}\ \emph {et~al.}(2003)\citenamefont {Agostinelli} \emph {et~al.}}]{GEANT4:2002zbu}%
  \BibitemOpen
  \bibfield  {author} {\bibinfo {author} {\bibfnamefont {S.}~\bibnamefont {Agostinelli}} \emph {et~al.} (\bibinfo {collaboration} {GEANT4}),\ }\href {https://doi.org/10.1016/S0168-9002(03)01368-8} {\bibfield  {journal} {\bibinfo  {journal} {Nucl. Instrum. Meth. A}\ }\textbf {\bibinfo {volume} {506}},\ \bibinfo {pages} {250} (\bibinfo {year} {2003})}\BibitemShut {NoStop}%
\bibitem [{\citenamefont {Abbasi}\ \emph {et~al.}()\citenamefont {Abbasi} \emph {et~al.}}]{icecube_phystools}%
  \BibitemOpen
  \bibfield  {author} {\bibinfo {author} {\bibfnamefont {R.}~\bibnamefont {Abbasi}} \emph {et~al.} (\bibinfo {collaboration} {IceCube Collaboration}),\ }\href@noop {} {\bibinfo {title} {{\texttt{PhysTools}}: Icecube physics utility library}},\ \bibinfo {howpublished} {\url{https://github.com/icecube/PhysTools}},\ \bibinfo {note} {accessed: July 2025}\BibitemShut {NoStop}%
\bibitem [{\citenamefont {Gumbel}(1958)}]{gumbel1958statistics}%
  \BibitemOpen
  \bibfield  {author} {\bibinfo {author} {\bibfnamefont {E.~J.}\ \bibnamefont {Gumbel}},\ }\href@noop {} {\emph {\bibinfo {title} {Statistics of Extremes}}}\ (\bibinfo  {publisher} {Columbia University Press},\ \bibinfo {year} {1958})\BibitemShut {NoStop}%
\bibitem [{\citenamefont {Arg\"uelles}\ \emph {et~al.}(2019)\citenamefont {Arg\"uelles}, \citenamefont {Schneider},\ and\ \citenamefont {Yuan}}]{Arguelles:2019izp}%
  \BibitemOpen
  \bibfield  {author} {\bibinfo {author} {\bibfnamefont {C.~A.}\ \bibnamefont {Arg\"uelles}}, \bibinfo {author} {\bibfnamefont {A.}~\bibnamefont {Schneider}},\ and\ \bibinfo {author} {\bibfnamefont {T.}~\bibnamefont {Yuan}},\ }\href {https://doi.org/10.1007/JHEP06(2019)030} {\bibfield  {journal} {\bibinfo  {journal} {JHEP}\ }\textbf {\bibinfo {volume} {06}},\ \bibinfo {pages} {030}},\ \Eprint {https://arxiv.org/abs/1901.04645} {arXiv:1901.04645 [physics.data-an]} \BibitemShut {NoStop}%
\bibitem [{\citenamefont {Wilks}(1938)}]{Wilks:1938dza}%
  \BibitemOpen
  \bibfield  {author} {\bibinfo {author} {\bibfnamefont {S.~S.}\ \bibnamefont {Wilks}},\ }\href {https://doi.org/10.1214/aoms/1177732360} {\bibfield  {journal} {\bibinfo  {journal} {Annals Math. Statist.}\ }\textbf {\bibinfo {volume} {9}},\ \bibinfo {pages} {60} (\bibinfo {year} {1938})}\BibitemShut {NoStop}%
\bibitem [{\citenamefont {Neumeier}\ \emph {et~al.}(2012)\citenamefont {Neumeier}, \citenamefont {Hofmann}, \citenamefont {Oberauer}, \citenamefont {Potzel}, \citenamefont {Schonert}, \citenamefont {Dandl}, \citenamefont {Heindl}, \citenamefont {Ulrich},\ and\ \citenamefont {Wieser}}]{Neumeier:2012cz}%
  \BibitemOpen
  \bibfield  {author} {\bibinfo {author} {\bibfnamefont {A.}~\bibnamefont {Neumeier}}, \bibinfo {author} {\bibfnamefont {M.}~\bibnamefont {Hofmann}}, \bibinfo {author} {\bibfnamefont {L.}~\bibnamefont {Oberauer}}, \bibinfo {author} {\bibfnamefont {W.}~\bibnamefont {Potzel}}, \bibinfo {author} {\bibfnamefont {S.}~\bibnamefont {Schonert}}, \bibinfo {author} {\bibfnamefont {T.}~\bibnamefont {Dandl}}, \bibinfo {author} {\bibfnamefont {T.}~\bibnamefont {Heindl}}, \bibinfo {author} {\bibfnamefont {A.}~\bibnamefont {Ulrich}},\ and\ \bibinfo {author} {\bibfnamefont {J.}~\bibnamefont {Wieser}},\ }\href {https://doi.org/10.1140/epjc/s10052-012-2190-z} {\bibfield  {journal} {\bibinfo  {journal} {Eur. Phys. J. C}\ }\textbf {\bibinfo {volume} {72}},\ \bibinfo {pages} {2190} (\bibinfo {year} {2012})},\ \Eprint {https://arxiv.org/abs/1511.07724} {arXiv:1511.07724 [physics.ins-det]} \BibitemShut {NoStop}%
\bibitem [{\citenamefont {Ishida}\ \emph {et~al.}(1997)\citenamefont {Ishida}, \citenamefont {Chen}, \citenamefont {Doke}, \citenamefont {Hasuike}, \citenamefont {Hitachi}, \citenamefont {Gaudreau}, \citenamefont {Kase}, \citenamefont {Kawada}, \citenamefont {Kikuchi}, \citenamefont {Komiyama}, \citenamefont {Kuwahara}, \citenamefont {Masuda}, \citenamefont {Okada}, \citenamefont {Qu}, \citenamefont {Suzuki},\ and\ \citenamefont {Takahashi}}]{ISHIDA1997380}%
  \BibitemOpen
  \bibfield  {author} {\bibinfo {author} {\bibfnamefont {N.}~\bibnamefont {Ishida}}, \bibinfo {author} {\bibfnamefont {M.}~\bibnamefont {Chen}}, \bibinfo {author} {\bibfnamefont {T.}~\bibnamefont {Doke}}, \bibinfo {author} {\bibfnamefont {K.}~\bibnamefont {Hasuike}}, \bibinfo {author} {\bibfnamefont {A.}~\bibnamefont {Hitachi}}, \bibinfo {author} {\bibfnamefont {M.}~\bibnamefont {Gaudreau}}, \bibinfo {author} {\bibfnamefont {M.}~\bibnamefont {Kase}}, \bibinfo {author} {\bibfnamefont {Y.}~\bibnamefont {Kawada}}, \bibinfo {author} {\bibfnamefont {J.}~\bibnamefont {Kikuchi}}, \bibinfo {author} {\bibfnamefont {T.}~\bibnamefont {Komiyama}}, \bibinfo {author} {\bibfnamefont {K.}~\bibnamefont {Kuwahara}}, \bibinfo {author} {\bibfnamefont {K.}~\bibnamefont {Masuda}}, \bibinfo {author} {\bibfnamefont {H.}~\bibnamefont {Okada}}, \bibinfo {author} {\bibfnamefont {Y.~H.}\ \bibnamefont {Qu}}, \bibinfo {author} {\bibfnamefont {M.}~\bibnamefont {Suzuki}},\ and\ \bibinfo {author} {\bibfnamefont {T.}~\bibnamefont
  {Takahashi}},\ }\href {https://doi.org/10.1016/S0168-9002(96)00740-1} {\bibfield  {journal} {\bibinfo  {journal} {Nuclear Instruments and Methods in Physics Research Section A: Accelerators, Spectrometers, Detectors and Associated Equipment}\ }\textbf {\bibinfo {volume} {384}},\ \bibinfo {pages} {380} (\bibinfo {year} {1997})}\BibitemShut {NoStop}%
\bibitem [{\citenamefont {Calvo}\ \emph {et~al.}(2018)\citenamefont {Calvo} \emph {et~al.}}]{ArDM:2016jbw}%
  \BibitemOpen
  \bibfield  {author} {\bibinfo {author} {\bibfnamefont {J.}~\bibnamefont {Calvo}} \emph {et~al.} (\bibinfo {collaboration} {ArDM}),\ }\href {https://doi.org/10.1016/j.astropartphys.2017.11.009} {\bibfield  {journal} {\bibinfo  {journal} {Astropart. Phys.}\ }\textbf {\bibinfo {volume} {97}},\ \bibinfo {pages} {186} (\bibinfo {year} {2018})},\ \Eprint {https://arxiv.org/abs/1611.02481} {arXiv:1611.02481 [astro-ph.IM]} \BibitemShut {NoStop}%
\bibitem [{\citenamefont {Neumeier}\ \emph {et~al.}(2015)\citenamefont {Neumeier}, \citenamefont {Dandl}, \citenamefont {Himpsl}, \citenamefont {Oberauer}, \citenamefont {Potzel}, \citenamefont {Sch\"onert},\ and\ \citenamefont {Ulrich}}]{Neumeier:2015lka}%
  \BibitemOpen
  \bibfield  {author} {\bibinfo {author} {\bibfnamefont {A.}~\bibnamefont {Neumeier}}, \bibinfo {author} {\bibfnamefont {T.}~\bibnamefont {Dandl}}, \bibinfo {author} {\bibfnamefont {A.}~\bibnamefont {Himpsl}}, \bibinfo {author} {\bibfnamefont {L.}~\bibnamefont {Oberauer}}, \bibinfo {author} {\bibfnamefont {W.}~\bibnamefont {Potzel}}, \bibinfo {author} {\bibfnamefont {S.}~\bibnamefont {Sch\"onert}},\ and\ \bibinfo {author} {\bibfnamefont {A.}~\bibnamefont {Ulrich}},\ }\href {https://doi.org/10.1209/0295-5075/111/12001} {\bibfield  {journal} {\bibinfo  {journal} {EPL}\ }\textbf {\bibinfo {volume} {111}},\ \bibinfo {pages} {12001} (\bibinfo {year} {2015})},\ \Eprint {https://arxiv.org/abs/1511.07725} {arXiv:1511.07725 [physics.ins-det]} \BibitemShut {NoStop}%
\bibitem [{\citenamefont {Fields}\ \emph {et~al.}(2023)\citenamefont {Fields}, \citenamefont {Gold}, \citenamefont {M.}, \citenamefont {McFadden}, \citenamefont {Elliott},\ and\ \citenamefont {Massarczyk}}]{Fields:2020wge}%
  \BibitemOpen
  \bibfield  {author} {\bibinfo {author} {\bibfnamefont {D.~E.}\ \bibnamefont {Fields}}, \bibinfo {author} {\bibnamefont {Gold}}, \bibinfo {author} {\bibfnamefont {R.~G.}\ \bibnamefont {M.}}, \bibinfo {author} {\bibfnamefont {N.}~\bibnamefont {McFadden}}, \bibinfo {author} {\bibfnamefont {S.~R.}\ \bibnamefont {Elliott}},\ and\ \bibinfo {author} {\bibfnamefont {R.}~\bibnamefont {Massarczyk}},\ }\href {https://doi.org/10.1016/j.nima.2022.167707} {\bibfield  {journal} {\bibinfo  {journal} {Nucl. Instrum. Meth. A}\ }\textbf {\bibinfo {volume} {1046}},\ \bibinfo {pages} {167707} (\bibinfo {year} {2023})},\ \Eprint {https://arxiv.org/abs/2009.10755} {arXiv:2009.10755 [physics.ins-det]} \BibitemShut {NoStop}%
\bibitem [{\citenamefont {Birks}(1951)}]{Birks:1951boa}%
  \BibitemOpen
  \bibfield  {author} {\bibinfo {author} {\bibfnamefont {J.~B.}\ \bibnamefont {Birks}},\ }\href {https://doi.org/10.1088/0370-1298/64/10/303} {\bibfield  {journal} {\bibinfo  {journal} {Proc. Phys. Soc. A}\ }\textbf {\bibinfo {volume} {64}},\ \bibinfo {pages} {874} (\bibinfo {year} {1951})}\BibitemShut {NoStop}%
\bibitem [{\citenamefont {Amoruso}\ \emph {et~al.}(2004)\citenamefont {Amoruso} \emph {et~al.}}]{AMORUSO2004275}%
  \BibitemOpen
  \bibfield  {author} {\bibinfo {author} {\bibfnamefont {S.}~\bibnamefont {Amoruso}} \emph {et~al.},\ }\href {https://doi.org/https://doi.org/10.1016/j.nima.2003.11.423} {\bibfield  {journal} {\bibinfo  {journal} {Nuclear Instruments and Methods in Physics Research Section A: Accelerators, Spectrometers, Detectors and Associated Equipment}\ }\textbf {\bibinfo {volume} {523}},\ \bibinfo {pages} {275} (\bibinfo {year} {2004})}\BibitemShut {NoStop}%
\bibitem [{\citenamefont {et~al. (Particle Data~Group)}(2024)}]{PDG2024}%
  \BibitemOpen
  \bibfield  {author} {\bibinfo {author} {\bibfnamefont {P.~Z.}\ \bibnamefont {et~al. (Particle Data~Group)}},\ }\href {https://doi.org/10.1093/ptep/ptae052} {\bibfield  {journal} {\bibinfo  {journal} {Prog. Theor. Exp. Phys.}\ }\textbf {\bibinfo {volume} {2024}},\ \bibinfo {pages} {083C01} (\bibinfo {year} {2024})}\BibitemShut {NoStop}%
\bibitem [{\citenamefont {Landau}\ and\ \citenamefont {Lifshitz}(1984)}]{landau1984electrodynamics}%
  \BibitemOpen
  \bibfield  {author} {\bibinfo {author} {\bibfnamefont {L.~D.}\ \bibnamefont {Landau}}\ and\ \bibinfo {author} {\bibfnamefont {E.~M.}\ \bibnamefont {Lifshitz}},\ }in\ \href@noop {} {\emph {\bibinfo {booktitle} {Electrodynamics of Continuous Media}}},\ \bibinfo {series} {Course of Theoretical Physics}, Vol.~\bibinfo {volume} {8}\ (\bibinfo  {publisher} {Pergamon Press},\ \bibinfo {address} {Amsterdam, The Netherlands},\ \bibinfo {year} {1984})\ \bibinfo {edition} {2nd}\ ed.,\ Chap.~\bibinfo {chapter} {XV}, pp.\ \bibinfo {pages} {394--412}\BibitemShut {NoStop}%
\bibitem [{\citenamefont {Seidel}\ \emph {et~al.}(2002)\citenamefont {Seidel}, \citenamefont {Lanou},\ and\ \citenamefont {Yao}}]{Seidel:2001vf}%
  \BibitemOpen
  \bibfield  {author} {\bibinfo {author} {\bibfnamefont {G.~M.}\ \bibnamefont {Seidel}}, \bibinfo {author} {\bibfnamefont {R.~E.}\ \bibnamefont {Lanou}},\ and\ \bibinfo {author} {\bibfnamefont {W.}~\bibnamefont {Yao}},\ }\href {https://doi.org/10.1016/S0168-9002(02)00890-2} {\bibfield  {journal} {\bibinfo  {journal} {Nucl. Instrum. Meth. A}\ }\textbf {\bibinfo {volume} {489}},\ \bibinfo {pages} {189} (\bibinfo {year} {2002})},\ \Eprint {https://arxiv.org/abs/hep-ex/0111054} {arXiv:hep-ex/0111054} \BibitemShut {NoStop}%
\bibitem [{\citenamefont {Mie}(1908)}]{Mie1908}%
  \BibitemOpen
  \bibfield  {author} {\bibinfo {author} {\bibfnamefont {G.}~\bibnamefont {Mie}},\ }\href {https://doi.org/10.1002/andp.19083300302} {\bibfield  {journal} {\bibinfo  {journal} {Annalen der Physik}\ }\textbf {\bibinfo {volume} {330}},\ \bibinfo {pages} {377} (\bibinfo {year} {1908})}\BibitemShut {NoStop}%
\bibitem [{\citenamefont {Asaadi}\ \emph {et~al.}(2019)\citenamefont {Asaadi}, \citenamefont {Jones}, \citenamefont {Tripathi}, \citenamefont {Parmaksiz}, \citenamefont {Sullivan},\ and\ \citenamefont {Williams}}]{Asaadi:2018ixs}%
  \BibitemOpen
  \bibfield  {author} {\bibinfo {author} {\bibfnamefont {J.}~\bibnamefont {Asaadi}}, \bibinfo {author} {\bibfnamefont {B.~J.~P.}\ \bibnamefont {Jones}}, \bibinfo {author} {\bibfnamefont {A.}~\bibnamefont {Tripathi}}, \bibinfo {author} {\bibfnamefont {I.}~\bibnamefont {Parmaksiz}}, \bibinfo {author} {\bibfnamefont {H.}~\bibnamefont {Sullivan}},\ and\ \bibinfo {author} {\bibfnamefont {Z.~G.~R.}\ \bibnamefont {Williams}},\ }\href {https://doi.org/10.1088/1748-0221/14/02/P02021} {\bibfield  {journal} {\bibinfo  {journal} {JINST}\ }\textbf {\bibinfo {volume} {14}}\bibfield  {number} {\bibinfo  {number} { (02)},\ \bibinfo {pages} {P02021}},\ }\Eprint {https://arxiv.org/abs/1804.00011} {arXiv:1804.00011 [physics.ins-det]} \BibitemShut {NoStop}%
\bibitem [{\citenamefont {Ignarra}(2014)}]{Ignarra:2014yqa}%
  \BibitemOpen
  \bibfield  {author} {\bibinfo {author} {\bibfnamefont {C.~M.}\ \bibnamefont {Ignarra}},\ }\emph {\bibinfo {title} {{Sterile Neutrino Searches in MiniBooNE and MicroBooNE}}},\ \href {https://doi.org/10.2172/1248362} {Ph.D. thesis},\ \bibinfo  {school} {MIT, Cambridge, Dept. Phys.} (\bibinfo {year} {2014})\BibitemShut {NoStop}%
\bibitem [{\citenamefont {Brice}\ \emph {et~al.}(2006)\citenamefont {Brice} \emph {et~al.}}]{MiniBooNE:2006fhd}%
  \BibitemOpen
  \bibfield  {author} {\bibinfo {author} {\bibfnamefont {S.~J.}\ \bibnamefont {Brice}} \emph {et~al.} (\bibinfo {collaboration} {MiniBooNE}),\ }\href {https://doi.org/10.1016/j.nima.2006.02.180} {\bibfield  {journal} {\bibinfo  {journal} {Nucl. Instrum. Meth. A}\ }\textbf {\bibinfo {volume} {562}},\ \bibinfo {pages} {97} (\bibinfo {year} {2006})},\ \Eprint {https://arxiv.org/abs/1005.3525} {arXiv:1005.3525 [physics.ins-det]} \BibitemShut {NoStop}%
\bibitem [{\citenamefont {Amaudruz}\ \emph {et~al.}(2019)\citenamefont {Amaudruz} \emph {et~al.}}]{DEAP:2017fgw}%
  \BibitemOpen
  \bibfield  {author} {\bibinfo {author} {\bibfnamefont {P.~A.}\ \bibnamefont {Amaudruz}} \emph {et~al.} (\bibinfo {collaboration} {DEAP}),\ }\href {https://doi.org/10.1016/j.nima.2018.12.058} {\bibfield  {journal} {\bibinfo  {journal} {Nucl. Instrum. Meth. A}\ }\textbf {\bibinfo {volume} {922}},\ \bibinfo {pages} {373} (\bibinfo {year} {2019})},\ \Eprint {https://arxiv.org/abs/1705.10183} {arXiv:1705.10183 [physics.ins-det]} \BibitemShut {NoStop}%
\bibitem [{\citenamefont {Abbasi}\ \emph {et~al.}(2010)\citenamefont {Abbasi} \emph {et~al.}}]{Abbasi_2010}%
  \BibitemOpen
  \bibfield  {author} {\bibinfo {author} {\bibfnamefont {R.}~\bibnamefont {Abbasi}} \emph {et~al.},\ }\href {https://doi.org/10.1016/j.nima.2010.03.102} {\bibfield  {journal} {\bibinfo  {journal} {Nuclear Instruments and Methods in Physics Research Section A: Accelerators, Spectrometers, Detectors and Associated Equipment}\ }\textbf {\bibinfo {volume} {618}},\ \bibinfo {pages} {139–152} (\bibinfo {year} {2010})}\BibitemShut {NoStop}%
\bibitem [{\citenamefont {Brigatti}\ \emph {et~al.}(2005)\citenamefont {Brigatti}, \citenamefont {Ianni}, \citenamefont {Lombardi}, \citenamefont {Ranucci},\ and\ \citenamefont {Smirnov}}]{Brigatti_2005}%
  \BibitemOpen
  \bibfield  {author} {\bibinfo {author} {\bibfnamefont {A.}~\bibnamefont {Brigatti}}, \bibinfo {author} {\bibfnamefont {A.}~\bibnamefont {Ianni}}, \bibinfo {author} {\bibfnamefont {P.}~\bibnamefont {Lombardi}}, \bibinfo {author} {\bibfnamefont {G.}~\bibnamefont {Ranucci}},\ and\ \bibinfo {author} {\bibfnamefont {O.}~\bibnamefont {Smirnov}},\ }\href {https://doi.org/10.1016/j.nima.2004.07.248} {\bibfield  {journal} {\bibinfo  {journal} {Nuclear Instruments and Methods in Physics Research Section A: Accelerators, Spectrometers, Detectors and Associated Equipment}\ }\textbf {\bibinfo {volume} {537}},\ \bibinfo {pages} {521–536} (\bibinfo {year} {2005})}\BibitemShut {NoStop}%
\end{thebibliography}%

\end{document}